\documentclass[11pt]{article}

\usepackage[english]{babel}
\usepackage{amsmath,amsthm}

\usepackage{latexsym} 
\usepackage{amsfonts}
\usepackage{amsmath}
\usepackage{amsthm}
\usepackage{amssymb}
\renewcommand{\theequation}{\thesection.\arabic{equation}}
\newcounter{subequation}[equation]
\makeatletter

\expandafter\let\expandafter\reset@font\csname reset@font\endcsname

\def\subeqnarray{\arraycolsep1pt
	\def\@eqnnum\stepcounter##1{\stepcounter{subequation}%
		{\reset@font\rm(\theequation\alph{subequation})}}
	\jot5mm     \eqnarray}

\makeatother

\def\str{\mathop{\hbox{\rm str}}\nolimits}

\def\be{\begin{equation}}
\def\ee{\end{equation}}

\def\lb{\label}

\def\bea{\begin{eqnarray}}
\def\eea{\end{eqnarray}}
\def\ba{\begin{array}}
	\def\ea{\end{array}}
\def\dd{\partial}
\def\o{\omega}

\def\one#1{#1^{\raise5pt\hbox{$\scriptstyle\!\!\!\!1$}}\,{}}
\def\two#1{#1^{\raise5pt\hbox{$\scriptstyle\!\!\!\!2$}}\,{}}

\def\tilde{\widetilde}
\def\II{\hbox{{1}\kern-.25em\hbox{l}}}
\def\a{\alpha}
\def\b{\beta}

\def\e{\varepsilon}
\def\qed{\rule{5pt}{5pt}}
\makeatletter
\def\binrel@#1{\begingroup
	\setboxz@h{\thinmuskip0mu
		\medmuskip\m@ne mu\thickmuskip\@ne mu
		\setbox\tw@\hbox{$#1\m@th$}\kern-\wd\tw@
		${}#1{}\m@th$}%
	\edef\@tempa{\endgroup\let\noexpand\binrel@@
		\ifdim\wdz@<\z@ \mathbin
		\else\ifdim\wdz@>\z@ \mathrel
		\else \relax\fi\fi}%
	\@tempa
}
\let\binrel@@\relax
\def\overset#1#2{\binrel@{#2}%
	\binrel@@{\mathop{\kern\z@#2}\limits^{#1}}}
\def\underset#1#2{\binrel@{#2}%
	\binrel@@{\mathop{\kern\z@#2}\limits_{#1}}}
\makeatother
\newfont{\bbd}{msbm10 scaled\magstep1}

\newtheorem{proposition}{Proposition}

\begin{document}

	\begin{center}
		{\LARGE {Yang-Baxter $R$-operators \\ for $osp$ superalgebras}}

 \vspace{0.3cm}

\large \sf
A.P. Isaev$^{a,b,d}$\footnote{\sf e-mail:isaevap@theor.jinr.ru},
D. Karakhanyan$^{a,c}$\footnote{\sf e-mail: karakhan@yerphi.am}, 
 R. Kirschner$^e$\footnote{\sf e-mail:Roland.Kirschner@itp.uni-leipzig.de} \\

\vspace{0.5cm}

\begin{itemize}
\item[$^a$]
{\it Bogoliubov Laboratory of Theoretical Physics,
JINR, Dubna, Russia}
\item[$^b$]
{\it Physics Faculty, M.V.~Lomonosov State University, Moscow}
\item[$^c$]
{\it Yerevan Physics Institute,
 2 Alikhanyan br., 0036 Yerevan, Armenia}
 \item[$^d$]{\it St.Petersburg Department of
 Steklov Mathematical Institute of RAS,\\
Fontanka 27, 191023 St. Petersburg, Russia}
 \item[$^e$]{\it Institut f\"ur Theoretische
 Physik, Universit\"at Leipzig, \\
 PF 100 920, D-04009 Leipzig, Germany}
\end{itemize}
\end{center}
\vspace{0.5cm}
\begin{abstract}
\noindent
We study Yang-Baxter equations with orthosymplectic supersymmetry.
We extend a new approach of the construction of the
spinor and metaplectic $\hat{\cal R}$-operators with
orthogonal and symplectic symmetries to the supersymmetric case
of orthosymplectic symmetry. In this approach the
orthosymplectic $\hat{\cal R}$-operator is given by the ratio
 of two operator valued Euler Gamma-functions. We
illustrate this approach by calculating such
$\hat{\cal R}$ operators in
explicit form for special cases of the $osp(n|2m)$ algebra,
in particular for a few low-rank cases.
We also propose a novel,  simpler and more elegant,
 derivation of the Shankar-Witten type formula for the $osp$ invariant
 $\hat{\cal R}$-operator and
 demonstrate the equivalence of the previous
 approach to the new one in the general case
 of the $\hat{\cal R}$-operator invariant under
 the action of the $osp(n|2m)$ algebra.
\end{abstract}

\section{Introduction}
\setcounter{equation}0

The similarities between the orthosymplectic
supergroups $OSp(N|M)$ (here $M=2m$ is an even number)
and their orthogonal $SO(N)$ and symplectic $Sp(M)$
bosonic  subgroups can be traced back to the existence
of invariant metrics in the (super)spaces ${\cal V}_{(N|M)}$,
${\cal V}_{N}$ and ${\cal V}_{M}$ of their defining representations.
These similarities lead to the consideration
of the supergroup $OSp$ and its superalgebra $osp$
in full analogy with the unified
treatment  (see e.g. \cite{IKK15}) of the groups $SO$, $Sp$
and their Lie algebras. Moreover these
similarities are inherited in the study of
solutions of the Yang-Baxter equations that possess such symmetries.

In the present paper, we continue our study \cite{FIKK}
of the solutions of the Yang-Baxter equations symmetric with respect
to ortho-symplectic groups.
We start with the graded RLL-relations with the $R$-matrix
in the defining representation
$R \in {\rm End}({\cal V}_{(N|M)} \otimes {\cal V}_{(N|M)})$
and  find the $L$-operator, 
$L(u) \in  {\rm End }({\cal V}_{(N|M)}) \otimes {\cal A}$, where
 ${\cal A}$ is a super-oscillator algebra
 invariant under the action of the $OSp(N|M)$ group.
 Then this $L$ operator allows one
 (via another type of $RLL$ relations) to define a richer and more
complicated family of solutions of  the Yang-Baxter equations,
namely the $\hat{\mathcal{R}}$-operators, which take  values in
 in the tensor product ${\cal A} \otimes {\cal A}$
 and are expressed as an expansion over the invariants in  ${\cal A} \otimes {\cal A}$.
The orthogonal and symplectic groups are embedded in the ortho-symplectic  super-group $OSp$,
and the $\hat{\mathcal{R}}$-operators invariant under the
$so(N)$ and $sp(M)$ algebras
can be obtained from the $OSp$-invariant
$\hat{\mathcal{R}}$-operator as special cases.
In the orthogonal case the
 algebra ${\cal A}$ is the $N$-dimensional Clifford algebra
 and the operator $\hat{\mathcal{R}}$ is called the spinor $R$-matrix.
 In the symplectic case the algebra ${\cal A}$ is the oscillator
 algebra and $\hat{\mathcal{R}}$  is called the
 metaplectic $R$-operator.

The standard approach to the problem of finding the spinor
($so$-invariant) $\hat{\cal R}$-operator was
developed in \cite{Witten}, \cite{CDI}
and is based on the expansion of the $\hat{\cal R}$-operator over the invariants
$I_k$ realized in the spaces ${\cal A} \otimes {\cal A}$.
Here the factors ${\cal A}$ are the Clifford algebras
 with the generators $(c^a)^\a_\b$, where $\a$, $\b$ and $a$
are respectively spinor  and vector indices. Then the invariants $I_k$ are
given by the contraction of the antisymmetrized products of
$c_1^{(a_1}\ldots c_1^{a_k)} \in {\cal A} \otimes I$ and
$c_2^{(b_1}\ldots c_2^{b_k)} \in I \otimes {\cal A}$
 with the invariant metrics $\varepsilon_{a_i b_i}$. In that approach
 we obtain the spinor $\hat{\cal R}$-operator as a sum over invariants
 $I_k$ with the coefficients $r_k$ which obey  recurrence relations.
 Analogous formulae of the Shankar-Witten (SW) type for the
 $\hat{\cal R}$-operators
 were deduced for the symplectic case in \cite{IKK15}
 and then were generalized for the ortho-symplectic case
 in \cite{FIKK}. Note that we cannot
consider these expressions for the $\hat{\cal R}$-operators as
quite satisfactory, since they do not provide closed formulas for
the considered $\hat{\cal R}$-operators. For example,
in the symplectic and ortho-symplectic cases, the sum over $I_k$
is infinite.

On the other
hand, it is known that an analogous $\hat{\cal R}$-operator invariant
under the $s\ell(2)$ algebra can be
represented (see \cite{FTT}, \cite{Fadd})
 in a compact form of the ratio
of two operator-valued Euler Gamma-functions.
Surprisingly, as it was shown in a recent
paper \cite{KaKi}, the $so$ and $sp$ invariant $\hat{\cal R}$-operators
(for special Clifford and oscillator representations of
$so$ and $sp$) are also represented in the Faddeev-Tarasov-Takhtajan
(FTT) form of the ratio of two operator-valued Euler Gamma-functions.

In the present paper, we generalize
the results of \cite{KaKi} to the supersymmetric case and
show that the $osp$ invariant $\hat{\cal R}$-operator
can also be represented in the FTT form. This is the
main result of our paper.
The natural conjecture is that the $osp$-invariant
SW type $\hat{\cal R}$-operator
given as a sum over invariants $I_k$ is equal to
the $osp$-invariant FTT type $\hat{\cal R}$-operator given by
the ratio of two Gamma-functions.
This conjecture is based on the fact that both
$\hat{\cal R}$-operators are solutions of the same system of
finite-difference equations which arise
from the RLL relations.

A complete proof of this
conjecture is still missing. In
the present paper we propose another simpler and more elegant
 derivation of the SW type formula for the $osp$ invariant
 $\hat{\cal R}$-operator.
  This new derivation supports the conjecture of
 the equivalence of the SW and FTT expressions for the
 $\hat{\cal R}$-operators.
Indeed, in the previous derivation, the role of invariant, "colorless",
elements in ${\cal A} \otimes {\cal A} $ is played by the operators $I_k$.
In the new derivation, we
prove that the invariants $I_k$ are polynomials of  one
invariant $I_1 \sim z$ only and
rewrite the RLL relation itself into a "colorless" form from the
very beginning in terms  of a system of
finite-difference equations in the variable $z$.

We relate this new system of equations to both the SW and the FTT
expressions for the $\hat{\cal R}$-operator. On one hand,  the FTT type
$\hat{\cal R}$-operator is its solution. On the other hand we show that
 the expansion of the SW type $\hat{\cal R}$-operator over $I_k$ satisfies
this system of finite-difference equations as well.

\vspace{0.2cm}

The paper is organized as follows. In Section 2, we
recall some basic facts of the linear algebra on the superspace
${\cal V}_{(N|M)}$ with $N$ bosonic and
$M$ fermionic coordinates and briefly formulate the theory of supergroups $OSp(N|M)$
and their Lie superalgebras $osp(N|M)$. In this section we fix
 our notation and conventions.
In Section 3, we define the $osp$-invariant solution of the Yang-Baxter
equation as an image of a special element of the Brauer algebra in the
tensor representation in super-spaces ${\cal V}_{(N|M)}^{\otimes r}$.
Section 4 is devoted to the formulation of the graded $RLL$ relations.
In this Section, we find a special $L$-operator that solves the
$RLL$ relations in the case of the $osp$ algebra and introduce
(see also \cite{FIKK}) the notion of the linear evaluation of the
Yangian ${\cal Y}(osp)$. In  Section 5 we define the
super-oscillator algebra ${\cal A}$ and describe the super-oscillator
representation for the linear evaluation of the Yangian ${\cal Y}(osp)$.
In particular, we define the set of $Osp$ invariant operators
$I_k$ in ${\cal A} \otimes {\cal A}$ and their generating function.

In terms of these
invariant operators we construct in Section 6 the $osp$ invariant
$\hat{\cal R}$-operators in the super-oscillator representation. We
find two forms for such $\hat{\cal R}$-operator. One of these forms
represents the $\hat{\cal R}$-operator as a ratio of Euler Gamma-functions.
For the $s\ell(2)$ case this type solution
was first obtained in \cite{FTT} (see also \cite{Fadd})
 and we call these solutions  the FTT type $\hat{\cal R}$-operators.
Another form of the $osp$ invariant $\hat{\cal R}$-operator in the super-
oscillator representation generalizes the SW solution \cite{Witten} of the
spinor-spinor $so$-invariant $\hat{\cal R}$-operator.
This solution (see eqs. (\ref{anzR}) and (\ref{sol-rm}))
for the $osp$-invariant $\hat{\cal R}$-matrix in the super-
oscillator representation was first obtained in our paper
\cite{FIKK} by using the methods developed
in \cite{CDI}, \cite{CDI2} and \cite{IKK15}.
 In \cite{FIKK} we have generalized
formulas for the $so$-type $R$-matrices
(in the Clifford
algebra representation) obtained in \cite{Witten}, \cite{CDI2} (see also
\cite{KarT},\cite{ZamL},\cite{Resh},\cite{Ogiev},\cite{IKK15}).
 In \cite{FIKK} we have
 also generalized the formulae for $sp$-type $R$-matrices
(in the oscillator, or metaplectic, representation of the Lie algebra $sp$),
which were deduced in \cite{IKK15}. It has been shown in \cite{FIKK} that
all these $so$- and $sp$-invariant $R$-matrices are obtained
from (\ref{anzR}), (\ref{sol-rm}) by restriction to the
corresponding bosonic Lie subalgebras of $osp$.

In Section 7 the result for the FTT type
 $R$ operator is studied in detail in particular
cases of $osp(N|M)$. The arguments of the Gamma-functions involve the
invariant operator $z \sim I_1$ which decomposes into a bosonic and a fermionic
part.
The finite spectral decomposition of the fermionic part is considered and
used to
decompose the $R$ operator with respect to the correspoding
projection operators.

In Section 8
we present the new and more direct derivation of the solutions
(\ref{anzR}) and (\ref{sol-rm}).
  Two Appendices are devoted to the proofs of the statements made in
 the main body of the paper.

\section{The ortho-symplectic supergroup and its Lie superalgebra}
\label{supgr}

 Consider (see, e.g., \cite{Ber}, \cite{FIKK}) a superspace ${\cal
 V}_{(N|M)}$
 with graded coordinates $z^a$ ($a=1,\dots,N+M$).
 The grading
 ${\rm grad}(z^a)$ of the coordinate~$z^a$ will be
 denoted as $[a]=0,1 \, ({\rm mod}2)$.
 If the coordinate $z^a$ is
  even then $[a]=0 \, ({\rm mod}2)$, and if the coordinate $z^a$ is
  odd then $[a]= 1 \, ({\rm mod}2)$.
  It means that the coordinates $z^a$ and $w^b$
  of two supervectors
 $z,w \in {\cal V}_{(N|M)}$ commute as follows
\begin{equation}
 \label{commzw}
 z^a \, w^b = (-1)^{[a]  [b]}  \, w^b \, z^a \; .
 \end{equation}
Let the superspace ${\cal V}_{(N|M)}$ be endowed with a bilinear form
\begin{equation}
\label{bf}
(z\cdot w) \equiv \varepsilon_{ab} z^a  w^b = z^a w_a =
  z_b w_a \bar{\varepsilon}^{ab} \; , \;\;\;\;
  (z\cdot w) = \epsilon (w \cdot z) \; ,
\end{equation}
which is symmetric for $\epsilon =+1$ and skewsymmetric for $\epsilon =-1$.
 In eq. (\ref{bf}) we define $w_a \equiv \varepsilon_{ab} w^b$,
where, in accordance with the last
relation in (\ref{bf}), the super-metric $\varepsilon_{ab}$
and inverse super-metric $\bar{\varepsilon}^{ab}$
have the properties
\begin{equation} \label{SupMet1}
 \varepsilon_{ab} \bar{\varepsilon}^{bd} =
 \bar{\varepsilon}^{db} \varepsilon_{ba} = \delta^d_a \; , \;\;\;\;\;
\varepsilon_{ab} = \epsilon (-1)^{[a][b]} \varepsilon_{ba} \;\;
\Leftrightarrow
\;\;
\bar{\varepsilon}^{ab} = \epsilon (-1)^{[a][b]} \bar{\varepsilon}^{ba}
 \; .
\end{equation}
 We stress that the super-metric $\varepsilon_{ab}$ is an even
 matrix in the sense that $\varepsilon_{ab}\neq 0$
 iff $[a]+[b]=0 \; ({\rm mod}2)$:
\begin{equation}\label{SupMet2}
\varepsilon_{ab}= (-1)^{[a]+[b]} \varepsilon_{ab} .
\end{equation}
 In other words the supermatrix
 $\varepsilon_{ab}$ is block-diagonal and its
  non-diagonal blocks vanish. Using (\ref{SupMet2}), the
 properties (\ref{SupMet1}) can be written as
 \begin{equation} \label{SupMet0}
 \varepsilon_{ab} = \epsilon (-1)^{[a]} \varepsilon_{ba} =
 \epsilon (-1)^{[b]} \varepsilon_{ba} \; , \;\;\;\;
 \bar{\varepsilon}^{ab} = \epsilon (-1)^{[a]} \bar{\varepsilon}^{ba} =
 \epsilon (-1)^{[b]} \bar{\varepsilon}^{ba} \; .
  \end{equation}

Further, we use
 the following agreement on raising and lowering indices
 for super-tensor components
 \begin{equation}
\label{agree}
{z^{...c}}\,_a^{\;\, d...} =
\varepsilon_{ab} \, z^{...c\, b\, d...} \; , \;\;\;
{z_{...c}^{\;\;\;\;\; a}}_{\, d...}
=  \bar{\varepsilon}^{ab} \, z_{...cbd...} \; .
\end{equation}
 According to this rule, we have
 $\varepsilon^{ab} = \bar{\varepsilon}^{ac}  \bar{\varepsilon}^{bd}
 \varepsilon_{cd} = \bar{\varepsilon}^{ba}$
 and the metric tensor with the upper indices $\varepsilon^{ab}$
 does not coincide with the inverse matrix $\bar{\varepsilon}^{ab}$.
 Further, we use only the inverse matrix $\bar{\varepsilon}^{ab}$
 and never the metric tensor $\varepsilon^{ab}$.

Consider a linear transformation in ${\cal V}_{(N|M)}$
\begin{equation}
\label{transU}
z^a \rightarrow z'^a=U^a_{\,\ b} \, z^b \; ,
\end{equation}
which preserves the
grading of the coordinates ${\rm grad}(z'^a) =  {\rm grad}(z^a)$. For the
elements $U^a_{\ b}$ of the supermatrix $U$
from (\ref{transU}) we have ${\rm grad}(U^a_{\ b}) =  [a] +[b]$.
The ortho-symplectic group $OSp$ is defined as the set of
supermatrices $U$ which
preserve the bilinear form (\ref{bf}) under the
transformations (\ref{transU})
\begin{equation} \label{OSpDef}
(-1)^{[c]([b]+[d])} \varepsilon_{ab} U^a_{\ c} U^b_{\ d} =  \varepsilon_{cd}
\;\; \Rightarrow \;\;
(-1)^{[c]([b]+[d])}  U^a_{\ c} U^b_{\ d} \bar{\varepsilon}^{cd} =
\bar{\varepsilon}^{ab} \; .
\end{equation}

Now we write the relations \eqref{OSpDef} in the coordinate-free form  as
\begin{equation} \label{OSpDef1}
\varepsilon_{\langle 12} \; U_1(-)^{12} U_2 (-)^{12} =
\varepsilon_{\langle 12}  \;\; \Leftrightarrow \;\;
U_1(-)^{12} U_2 (-)^{12} \; \bar{\varepsilon}^{12 \rangle} =
\bar{\varepsilon}^{12
\rangle} \; ,
\end{equation}
 where the concise matrix notation is used
 \begin{equation}
 \label{min12}
 \begin{array}{c}
 \bar{\varepsilon}^{12 \rangle} \in {\cal V}_{(N|M)}
 \otimes {\cal V}_{(N|M)}
 \; ,
 \;\;\;
  U_1 = U \otimes I \; , \;\;\;
  U_2 = I \otimes U  \; , \\ [0.3cm]
  ((-)^{12})^{a_1 a_2}_{\;\;\; b_1 b_2} = (-1)^{[a_1][a_2]}
  \delta^{a_1}_{\;\; b_1} \delta^{a_2}_{\;\; b_2} \; , \;\;\;
  (-)^{12} \in {\rm End}({\cal V}_{(N|M)} \otimes {\cal V}_{(N|M)}) \; .
  \end{array}
  \end{equation}
  Here $\otimes$ denotes the graded tensor product:
  $$
  (I \otimes B)(A \otimes I) = (-1)^{[A] \, [B]} \, (A \otimes B)
  \; , \;\;\;\;\;  (A \otimes I) (I \otimes B) =  (A \otimes B) \; ,
  $$
  and $[A] := {\rm grad}(A)$, $[B] := {\rm grad}(B)$.
  We remark that in our paper
  we use the convention in which gradation is carried by the
  coordinates, while there is another convention in which gradation
  is carried by the basis vectors (see e.g. \cite{Ragoucy}).
  The relation of these two formulations is explained in \cite{FIKK}.

Consider the elements $U \in OSp$ which are close to unity $I$:
$U  = I + A + \dots$. Here dots denote the terms which are much smaller
than $A$. In this case, the defining relations \eqref{OSpDef} give
conditions for the supermatrices $A$ which are interpreted as elements
 of the Lie superalgebra $osp$ of the supergroup $OSp$:
\begin{align}
 \label{osp}
(-1)^{[c]([b]+[d])} \varepsilon_{ab} ( \delta^a_{\ c} A^b_{\ d}  + A^a_{\ c}
\delta^b_d ) =
\left( (-1)^{[c]+[c][d]} \varepsilon_{cb} A^b_{\ d}
 + \varepsilon_{ad}  \, A^a_{\ c} \right) =0 \; ,
\end{align}
or equivalently
\begin{equation} \label{OspAlg}
A_{cd}=-\epsilon(-1)^{[c][d]+[c]+[d]} A_{dc} \; .
\end{equation}
The coordinate free form of relation (\ref{osp}) is
 \begin{equation} \label{OspAlg1}
 \varepsilon_{\langle 12} (A_1 + (-)^{12}A_2 (-)^{12})
 = 0 \;\;\;\; \Leftrightarrow \;\;\;\;
  (A_1 + (-)^{12}A_2 (-)^{12})  \bar{\varepsilon}^{12 \rangle} = 0 \; .
\end{equation}
One can directly deduce these relations
 from equalities (\ref{OSpDef1}).

 The set of super-matrices $A$,
 which satisfy (\ref{osp}), (\ref{OspAlg1}), forms a vector space over
 $\mathbb{C}$ which is denoted as $osp$. One can check that for two
 super-matrices
 $A,B \in osp$ the commutator
 \begin{equation}
 \label{osp05}
 [A,B]=AB - BA \; ,
 \end{equation}
 also obeys (\ref{osp}), (\ref{OspAlg1})
 and thus belongs to $osp$. It means that $osp$ is an algebra.
 Any matrix $A$ which satisfies (\ref{osp}), (\ref{OspAlg1}) can be
 represented
 as
\begin{equation}
 \label{osp02}
A^a_{\;\; c} = E^a_{\;\; c} -
 (-1)^{[c]+[c][d]} \varepsilon_{cb} E^b_{\;\; d} \bar{\varepsilon}^{da}
 \end{equation}
where $||E^a_{\;\; c}||$ is an arbitrary matrix. Let $\{e^{\;\; f}_{g} \}$ be
the matrix units, i.e., matrices with the components
 $(e^{\;\; f}_{g})^b_{\;\; d}= \delta^f_d \delta^b_g$.
 If we substitute $E = e^{f}_{\;\; g} = \bar{\varepsilon}^{f g'}
 \varepsilon_{g f'} e^{\;\;\; f'}_{g'}$ in (\ref{osp02}),
 then we obtain the basis elements $\{\tilde{G}^f_{\;\; g}\}$ in the space
 $osp$ of matrices (\ref{OspAlg1}):
 \begin{equation}
 \label{osp03}
 \begin{array}{c}
 (\tilde{G}^f_{\;\; g})^a_{\;\; c} \equiv (e^f_{\;\; g})^a_{\;\; c} -
 (-1)^{[c]+[c][d]} \varepsilon_{cb} (e^f_{\;\; g})^b_{\;\; d}
 \bar{\varepsilon}^{da}
 = \bar{\varepsilon}^{f a} \varepsilon_{g c} - \epsilon (-1)^{[c][a]}
 \delta^f_c \delta^a_g \; .
 \end{array}
\end{equation}
 Now any super-matrix $A \in osp$ which satisfies (\ref{osp}), (\ref{OspAlg1})
can be expanded over the basis  (\ref{osp03})
 \begin{equation}
 \label{osp11}
 A^a_{\;\; c} = a^g_{\;\; f} (\tilde{G}^f_{\;\; g})^a_{\;\; c} \; ,
 \end{equation}
 where $a^g_{\;\; f}$ are the components of the super-matrix.
Since the elements $(\tilde{G}^f_{\;\; g})^a_{\;\; c}$ are even, i.e.,
$(\tilde{G}^f_{\;\; g})^a_{\;\; c} \neq 0$ iff $[f]+[g]+[a]+[c] = 0 \; ({\rm mod}2)$,
then from the condition ${\rm grad}(A^a_{\;\; c}) =[a]+[c]$
we obtain that ${\rm grad}(a^g_{\;\; f}) =[g]+[f]$. It means that the usual commutator
(\ref{osp05}) appears as a super-commutator
 for the basis elements $\tilde{G}^f_{\;\; g}$:
 $$
 [A,B]^a_{\;\; c} = [ a^g_{\;\; f} (\tilde{G}^f_{\;\; g}) , \;
  b^n_{\;\; k} (\tilde{G}^k_{\;\; n})]^a_{\;\; c} =
 a^g_{\;\; f} b^n_{\;\; k}  \;
 \bigl([\tilde{G}^f_{\;\; g} , \;
 \tilde{G}^k_{\;\; n}]_{\pm} \bigr)^a_{\;\; c} \; ,
 $$
where in the component form the super-commutator is
 \begin{equation}
 \label{osp06b}
 \bigl( [\tilde{G}^{a_1}_{\;\; b_1} , \; \tilde{G}^{a_2}_{\;\; b_2}]_{\pm}
 \bigr)^{a_3}_{\;\; c_3}  \equiv
 (\tilde{G}^{a_1}_{\;\; b_1})^{a_3}_{\;\; b_3} \,
 (\tilde{G}^{a_2}_{\;\; b_2})^{b_3}_{\;\; c_3}
  - (-1)^{([a_1]+[b_1])([a_2]+[b_2])}
  (\tilde{G}^{a_2}_{\;\; b_2})^{a_3}_{\;\; b_3}
  (\tilde{G}^{a_1}_{\;\; b_1})^{b_3}_{\;\; c_3} \; .
  \end{equation}
  We notice that the elements of the matrices $\tilde{G}^{a}_{\;\; b}$
  are numbers. However, the super-commutator (\ref{osp06b})
  is written for $\tilde{G}^{a}_{\;\; b}$ as for the graded
  elements with ${\rm deg}(\tilde{G}^{a}_{\;\; b})= [a]+[b]$.

  Now we substitute the explicit representation (\ref{osp03})
  in the right-hand side of (\ref{osp06b}) and deduce the
  defining relations for the basis elements of the superalgebra $osp$:
 \begin{equation}
 \label{osp06}
 \begin{array}{c}
 (-1)^{[b_1][a_2]} \cdot
  [\tilde{G}^{a_1}_{\;\; b_1} , \; \tilde{G}^{a_2}_{\;\; b_2}]_{\pm}
  = -(-1)^{[a_1][a_2]} \, \bar{\varepsilon}^{a_1a_2} \, \tilde{G}_{b_1b_2} +
  \epsilon \, \delta^{a_2}_{b_1} \, \tilde{G}^{a_1}_{\;\;\; b_2} + \\ [0.3cm]
  + (-1)^{[a_1][a_2]}\, \varepsilon_{b_1b_2} \, \tilde{G}^{a_2a_1} -
  \epsilon (-1)^{[a_1]([b_1]+[a_2])} \, \delta^{a_1}_{b_2} \,
  \tilde{G}^{a_2}_{\;\;\; b_1} \; ,
 \end{array}
\end{equation}
where we have omitted the matrix indices. Below we use
 the standard component-free form of notation, where we substitute
 $(\tilde{G}^{a_i}_{\;\; b_i})^{a_k}_{\;\; b_k} \to \tilde{G}_{ik}$
 (here $i$ and $k$ are numbers $1,2,3$ of two super-spaces
 ${\cal V}_{(N|M)}$ in ${\cal V}_{(N|M)}^{\otimes 3}$). In this notation,
taking into account (\ref{osp06b}),
the relation (\ref{osp06}) is written as
  \begin{equation}
 \label{osp06a}
 [(-)^{12} \tilde{G}_{13} (-)^{12} \; , \; \tilde{G}_{23}] =
 [\epsilon {\cal P}_{12} - {\cal K}_{12}   \; , \; \tilde{G}_{23}] \; ,
\end{equation}
 where we introduce two matrices ${\cal K},{\cal P} \in
 {\rm End}({\cal V}_{(N|M)}^{\otimes 2})$:
\begin{equation}
 \label{osp07}
 {\cal K}^{a_1 a_2}_{b_1 b_2} = \bar{\varepsilon}^{a_1 a_2} \varepsilon_{b_1
 b_2}
 \; ,\; \;\;\;
 {\cal P}^{a_1 a_2}_{b_1 b_2} =
 (-1)^{[a_1][a_2]} \delta^{a_1}_{b_2} \delta^{a_2}_{b_1} \; .
\end{equation}
The matrix ${\cal P}$ is called superpermutation
 since it permutes super-spaces, e.g.,
using this matrix one can write (\ref{commzw}) as
 ${\cal P}^{ab}_{cd}w^c z^d = z^a w^b$. Note that
 the generators (\ref{osp03}) of the Lie super-algebra $osp$
 can be expressed in terms of ${\cal P}$ and ${\cal K}$ as
 \begin{equation}
 \label{GKP}
 \tilde{G} = {\cal K} - \epsilon \, {\cal P} \; ,
 \end{equation}
 and after  substituting
  (\ref{GKP}) into (\ref{osp06a})
  can be written (\ref{osp06a}) in the form
   \begin{equation}
 \label{osp06d}
 [(-)^{12} \tilde{G}_{13} (-)^{12} \; , \; \tilde{G}_{23}] +
 [\tilde{G}_{12}   \; , \; \tilde{G}_{23}] = 0 \; .
\end{equation}
One can explicitly check the relation (\ref{osp06d}) by making use of the
identities for the operators ${\cal P}$ and ${\cal K}$ presented in
 Appendix {\bf \ref{PKprop}}.

 Note that conditions (\ref{OspAlg1}) for the $osp$  generators
 $A^{a}_{\;\; c} = (\tilde{G}^{f}_{\;\; g})^{a}_{\;\; c}$,
 given in  (\ref{osp03}) and (\ref{GKP}), can be written as
 \begin{equation}
 \label{osp08b}
 {\cal K}_{12} (\tilde{G}_{31} + (-)^{12} \tilde{G}_{32} (-)^{12}) = 0 \; ,
 \;\;\;
(\tilde{G}_{31} + (-)^{12} \tilde{G}_{32} (-)^{12})  {\cal K}_{12}  = 0 \; .
 \end{equation}
 One can verify that these conditions are equivalent to
   \begin{equation}
 \label{osp08}
 {\cal K}_{12} ((-)^{12} \tilde{G}_{13}(-)^{12}  + \tilde{G}_{23} ) = 0 \; ,
 \;\;\;
((-)^{12} \tilde{G}_{13}(-)^{12} + \tilde{G}_{23})
 {\cal K}_{12}  = 0 \; .
 \end{equation}
 Using (\ref{osp08}) and the commutation
 relations of super-permutation ${\cal P}$
 and generators $\tilde{G}$ (see appendix {\bf \ref{PKprop}})
 \begin{equation}
 \label{osp09}
 {\cal P}_{12} (-)^{12} \tilde{G}_{13} (-)^{12} =
  \tilde{G}_{23} {\cal P}_{12}  \; , \;\;\;
(-)^{12} \tilde{G}_{13}  (-)^{12}  {\cal P}_{12} =
 {\cal P}_{12} \tilde{G}_{23}  \; ,
 \end{equation}
 we write (\ref{osp06a}) as
 \begin{equation}
 \label{osp10}
  [(-)^{12} \tilde{G}_{13} (-)^{12} \;  , \; \tilde{G}_{23}]
   = [\epsilon {\cal P}_{12} - {\cal K}_{12}  \; , \;
   (-)^{12} \tilde{G}_{13} (-)^{12}] \; .
 \end{equation}
It means that the defining relations (\ref{osp06})
 can be written in many equivalent forms. At the end of this section
 we note that the matrix (\ref{GKP}) is the split Casimir operator
 for the Lie superalgebra $osp$ in the defining representation.

\section{The OSp-invariant R-matrix and the graded Yang--Baxter
equation\label{sec2}}
\setcounter{equation}0

 Consider the three $OSp$ invariant operators in ${\cal V}_{(N|M)}^{\otimes 2}$:
 the identity operator $\mathbf{1}$, the super-permutation operator
 $\mathcal{P}$
 and metric operator $\mathcal{K}$.
 According to definition (\ref{osp07}),
 the super-permutation $\mathcal{P}_{12}$ is a product of the usual
 permutation $P_{12}$ and the sign factor $(-)^{12}$,
\begin{equation}
 \label{PP12}
\mathcal{P}_{12} = (-)^{12} P_{12} \, ,
\qquad \text{or in components} \qquad
\mathcal{P}^{a_1a_2}_{b_1b_2}=(-1)^{[a_1][a_2]}\delta^{a_1}_{b_2}
\delta^{a_2}_{b_1} ,
\end{equation}
while the operator $\mathcal{K}_{12}$ is defined as
 \begin{equation}
 \label{KK12}
\mathcal{K}_{12} = \bar{\varepsilon}^{12 \rangle} \varepsilon_{\langle 12}
\, , \qquad
\text{or in components}
\qquad\mathcal{K}^{a_1a_2}_{b_1b_2}=\bar{\varepsilon}^{a_1a_2}
 \varepsilon_{b_1b_2}.
 \end{equation}
Their $OSp$ invariance means that (see (\ref{OSpDef1}))
  \begin{equation}
 \label{ospinv}
 \begin{array}{c}
 U_1 (-)^{12} U_2 (-)^{12} \mathcal{K}_{12} =
 \mathcal{K}_{12} U_1 (-)^{12} U_2 (-)^{12} \; , \\ [0.2cm]
 U_1 (-)^{12} U_2 (-)^{12} \mathcal{P}_{12} =
 \mathcal{P}_{12} U_1 (-)^{12} U_2 (-)^{12} \; .
 \end{array}
 \end{equation}
 In particular, it follows from these relations that the
 comultiplication for the supermatrices $U \in Osp(N|M)$
 has the graded form $\Delta(U)_{12} = U_1 (-)^{12} U_2 (-)^{12}$.
 In fact this comultiplication follows from the transformation
 (\ref{transU}) applied to the second rank
 tensor $z^{a_1}\cdot  z^{a_2}$.

Using the operators $\mathcal{P},\mathcal{K}$ one can construct
a set of operators $\{s_i,e_i|{i=1,\dots,n-1}\}$ in ${\cal
 V}_{(N|M)}^{\otimes n}$:
\begin{equation}
 \label{Brauer}
s_i=\epsilon \mathcal{P}_{i,i+1} \equiv
\epsilon I^{\otimes (i-1)} \otimes \mathcal{P} \otimes I^{\otimes (n-i-1)}
\,
,\;\;\;\; e_i=\mathcal{K}_{i,i+1} \equiv
I^{\otimes (i-1)} \otimes \mathcal{K} \otimes I^{\otimes (n-i-1)} \, ,
\end{equation}
which define the matrix representation $T$
 of the Brauer algebra $B_n(\omega)$ \cite{Brauer},
\cite{Wenz} with the
parameter
\begin{equation}
\label{oMN}
\omega=\varepsilon_{cd} \, \bar{\varepsilon}^{cd}
=\epsilon (N-M) \; .
\end{equation}
Recall that here $N$ and $M$ are the numbers of even and odd
coordinates, respectively. Indeed, one can check directly (see Appendix
{\bf \ref{PKprop}}) that the operators (\ref{Brauer}) satisfy the defining
relations for the generators of the Brauer algebra $B_n(\omega)$
 \begin{equation}
 \label{defBrauer1}
 \begin{array}{c}
s^2_i = 1 \; , \;\;\;  e^2_i = \omega e_i \; , \;\;\;
s_i \, e_i = e_i \, s_i = e_i\; , \;\;\; i = 1, . . . , n -1 , \\
s_i s_j = s_j s_i \; , \;\;\; e_i e_j = e_j e_i \; , \;\;\; s_i e_j = e_j
s_i
\; , \;\;\;  |i - j| > 1,
 \end{array}
\end{equation}
\begin{equation}
 \label{defBrauer2}
 \begin{array}{c}
s_i \, s_{i+1} \, s_i = s_{i+1} \, s_i \, s_{i+1} \; , \;\;\; e_i \, e_{i+1}
\,
e_i = e_i
\; , \;\;\;  e_{i+1} \, e_i \, e_{i+1} = e_{i+1} \; , \\
s_i \, e_{i+1} \, e_i =
 s_{i+1} \, e_i \; , \;\;\; e_{i+1}\, e_i\, s_{i+1} = e_{i+1} \, s_i \; ,
 \;\;\;
  i = 1, . . . , n-2\; .
\end{array}
\end{equation}
 This presentation of the Brauer algebra can be obtained in the
 special limit $q \to 1$ from the BMW algebra presentation \cite{BW}
 and it is used in many investigations (see, e.g.,
 \cite{Nazar}, \cite{IsMol10}, \cite{IsMoOg}, \cite{IsPod}).
 We stress that the matrix representation $T$ (\ref{Brauer})
 of the generators $s_i,e_i \in B_n(\omega)$
 acts in the space ${\cal V}_{(N|M)}^{\otimes n}$.

Let us consider the following linear combination of the
 unit element $\mathbf{1} \in B_n(\omega)$ and the
 generators $s_i,e_i \in B_n(\omega)$
 \begin{equation}
 \label{RBrauer}
\hat{\rho}_i(u) =u(u+\beta) \, s_i- (u+\beta) \mathbf{1}+ u \, e_i \;\; \in
\;\;
 B_n(\omega) \; ,
\end{equation}
where $u$ is a spectral parameter and
\begin{equation}
\beta=1-\frac{\omega}{2} \; .
\end{equation}
\noindent
\begin{proposition} \label{Prop11}
 $ $ (see \cite{IsMol10},\cite{FIKK}).
The element (\ref{RBrauer}) satisfies the Yang-Baxter equation
 \begin{equation}
 \label{YBEbr}
\hat{\rho}_i(u) \hat{\rho}_{i+1}(u + v) \hat{\rho}_i(v) =
\hat{\rho}_{i+1}(v) \hat{\rho}_{i}(u + v) \hat{\rho}_{i+1}(u)  \; ,
\end{equation}
and the unitarity condition
 
 $\hat{\rho}_i(u) \hat{\rho}_{i}(-u)  =
(u^2-1)(u^2 - \beta^2) \mathbf{1}$.
 
 \end{proposition}

The matrix representation
 $T$ (\ref{Brauer}) of the element (\ref{RBrauer}) is
 \begin{equation}
\label{RBrauer01}
\hat{R}(u) \equiv \epsilon \, T(\hat{\rho}(u)) =
u(u+\beta)\mathcal{P}-\epsilon(u+\beta)\mathbf{1}+\epsilon u\mathcal{K} \; .
\end{equation}
Here we suppress index $i$ for simplicity.
It follows from (\ref{YBEbr})
that $\hat{R}(u)$ satisfies the braid version of the Yang--Baxter equation
\begin{equation} \label{eq:YBEbraid}
\hat{R}_{12}(u-v) \hat{R}_{23}(u) \hat{R}_{12}(v) = \hat{R}_{23}(v)
\hat{R}_{12}(u) \hat{R}_{23}(u-v).
\end{equation}
Thus, in the supersymmetric case the braid version
(\ref{eq:YBEbraid}) of the Yang--Baxter equation  is
 the same as in the non supersymmetric case.
 Further we  use the following $R$-matrix
\begin{align}
 \label{Rmatri}
R(u)&={\cal P} \hat{R}(u)=
(u-\frac{\omega}{2}+1)(u\mathbf{1}-\epsilon\mathcal{P})+u\mathcal{K}
\notag\\
&= u(u+\beta) \mathbf{1} -\epsilon (u+\beta) \mathcal{P} +u \mathcal{K} \; ,
\end{align}
 which is the image of the elements \cite{IsMol10}:
 $$
 \rho_i(u) =
  u(u+\beta) \mathbf{1} - (u+\beta) \, s_i + u \,  e_i \;\; \in \;\;
 B_n(\omega) \; .
 $$
\begin{proposition}\label{propSR}
 The standard $R$-matrix
$R(u)=\mathcal{P}\hat{R}(u)$,
which was defined in (\ref{Rmatri}),
 satisfies the graded version  of the Yang--Baxter equation \cite{KulSkl}
\begin{equation} \label{YBEgr}
R_{12}(u-v)(-)^{12} R_{13}(u)(-)^{12} R_{23}(v) = R_{23}(v) (-)^{12}
R_{13}(u)
(-)^{12} R_{12}(u-v).
\end{equation}
\end{proposition}
\noindent
{\bf Proof.}
 Substituting  $\hat{R}_{ij}(u)=\mathcal{P}_{ij}R_{ij}(u) =
(-)^{ij} P_{ij}R_{ij}(u)$ into
(\ref{eq:YBEbraid}) and moving all usual
permutations $P_{ij}$ to the
left
we
write (\ref{eq:YBEbraid}) in the form
 \begin{equation} \label{YBEgr01}
R_{23}(u-v)(-)^{13} R_{13}(u)(-)^{12} R_{12}(v) = R_{12}(v) (-)^{13}
R_{13}(u)
(-)^{23} R_{23}(u-v).
\end{equation}
The matrix $R \in {\rm End}({\cal V}_{(N|M)}^{\; \otimes 2})$
is an even matrix since
the following condition holds
\begin{equation} \label{Reven1}
R^{i_1i_2}_{j_1j_2} \neq 0 \qquad\; \text{iff}\qquad\;
 [i_1]+[i_2]+[j_1]+[j_2]=0 \; ({\rm mod} 2) \; .
\end{equation}
This follows from the same property for the matrices $\mathbf{1},
\mathcal{P},\mathcal{K}$ which compose the operator $R(u)$.
Therefore, for arbitrary $k$ we have
\begin{equation} \label{eq:Rmat-}
R_{ij}  (-)^{ik}(-)^{jk} = (-)^{ik}(-)^{jk} R_{ij}  \; .
\end{equation}
where the operator $(-)^{ik}$ is defined in (\ref{min12})
($i$ and $k$ are numbers of only two super-spaces ${\cal V}_{(N|M)}$
 in the product ${\cal V}_{(N|M)}^{\otimes n}$ where
  the operator $(-)^{ik}$ acts nontrivially).
Using the property (\ref{eq:Rmat-}), one can convert (\ref{YBEgr01}) into
the form
\begin{equation}
\label{RRR}
\begin{array}{c}
R_{23}(u-v) (-)^{12} R_{13}(u) (-)^{12} R_{12}(v) 
 =  R_{12}(v) (-)^{12} R_{13}(u) (-)^{12} R_{23}(u-v) \; ,
\end{array}
\end{equation}
and after the change of the spectral parameters we obtain the graded
 version of Yang--Baxter equation \eqref{YBEgr}. \hfill \qed

 \vspace{0.2cm}

\noindent
{\bf Remark 1.} We stress that the sign operators $(-)^{12}$  in
\eqref{YBEgr} can be substituted by the operators $(-)^{23}$ by means of
 manipulations similar
 to (\ref{eq:Rmat-}). Moreover, if $R_{ij}(u)$ solves the Yang-Baxter
 equation \eqref{YBEgr}, then
 the twisted $R$-matrix $(-)^{ij}R_{ij}(u)(-)^{ij}$ is also a
 solution of \eqref{YBEgr}.

\noindent
{\bf Remark 2.} Eqs. (\ref{RBrauer01}), (\ref{Rmatri})
 give unified forms for solutions of the Yang-Baxter
 equations (\ref{eq:YBEbraid}), (\ref{YBEgr}) which
are invariant under the action of all Lie (super)groups $SO$, $Sp$ and
$OSp$.
Recall that for the $SO$ case the $R$-matrix (\ref{Rmatri})
  was found in \cite{Zam} and for the $Sp$ case it was indicated
  in \cite{Karow}. For the $OSp$ case such $R$-matrices were considered in
 many papers (see, e.g., \cite{Kulish}, \cite{Ragoucy}, \cite{Isae}).

\section{Graded RLL-relation and the linear evaluation
of Yangian ${\cal Y}(osp)$}
\setcounter{equation}0

We start with the following graded form of the RLL-relation
(see, e.g., \cite{Isae} and references therein)
\begin{equation} \label{eq:RLL}
R_{12}(u-v)L_1(u)(-)^{12} L_2(v) (-)^{12} = (-)^{12} L_2(v) (-)^{12} L_1(u)
R_{12}(u-v) \; ,
\end{equation}
where the R-matrix is given in \eqref{Rmatri}. This graded
form of the $RLL$ relations
is also motivated by the invariance conditions (\ref{ospinv}).
It is known
 (see, e.g., \cite{FIKK}, \cite{Ragoucy} and references therein)
that eqs. (\ref{eq:RLL}) with the R-matrix \eqref{Rmatri}
 are defining relations for the super-Yangian ${\cal Y}(osp)$.
In  \cite{FIKK} we proved the following statement.
\begin{proposition}\label{prop2a} The $L$-operator
 \begin{equation} \label{L01}
L^a_{\;\; b}(u) = (u + \alpha) \, {\bf 1} \delta^a_b
 + G^a_{\;\; b} \; ,
 \end{equation}
where $\alpha$ is an arbitrary constant,
 solves the RLL-relation (\ref{eq:RLL}) iff $G^a_{\;\; b}$
 is a traceless matrix of generators of the Lie
 superalgebra $osp$, i.e.,  it satisfies
 equations  (cf. (\ref{osp08b}))
\begin{equation}
\label{SymCond}
\mathcal{K}_{12} \Bigl\{G_1  + (-)^{12} G_2 (-)^{12} \Bigr\}= 0
= \Bigl\{G_1  + (-)^{12} G_2 (-)^{12}   \Bigr\} \mathcal{K}_{12} \; ,
\end{equation}
 defining relations for $osp$-algebra (cf. (\ref{osp06a}))
 \begin{equation} \label{CommGb}
 G_1 (-)^{12} G_2 (-)^{12} - (-)^{12} G_2 (-)^{12} G_1   =
[\mathcal{K}_{12} - \epsilon\mathcal{P}_{12}, \; G_1 ] \; ,
\end{equation}
 and in addition  obeys the quadratic characteristic identity
\begin{equation} \label{Cond1}
G^2 + \beta \, G -\frac{\epsilon}{\omega} \,
 \mathrm{str}\left(G^2\right) \mathbf{1} =0 \; ,
\end{equation}
 where as usual $\beta = 1-\omega/2$.
\end{proposition}
\noindent
The $L$-operator (\ref{L01}), where the elements $G^a_{\; b}$ satisfy
the conditions (\ref{SymCond}), (\ref{CommG})
 and (\ref{Cond1}), is called {\em the linear evaluation} of the
 Yangian ${\cal Y}(osp)$.

\noindent
{\bf Remark 3.} The relations (\ref{CommGb}) are written after the
exchange $1 \leftrightarrow 2$ in the form
 \begin{equation} \label{CommG}
(-)^{12} G_1 (-)^{12} G_2  -  G_2 (-)^{12} G_1 (-)^{12}  =
[\epsilon\mathcal{P}_{12}-\tilde{\mathcal{K}}_{12}, \; G_2 ] \; ,
\end{equation}
where $\tilde{\cal K}_{12}= {\cal K}_{21}=
(-)^{12} {\cal K}_{12} (-)^{12}$,
or $\tilde{\cal K}^{a_1 a_2}_{\;\; b_1 b_2}=
 \bar{\varepsilon}^{a_2 a_1}\varepsilon_{b_2 b_1}$.
Now we are able to compare the defining relations (\ref{CommGb}),
(\ref{CommG})
with (\ref{osp06a}), (\ref{osp10}), where the elements $G^a_{\; b}$ are
represented as matrices $\tilde{G}^a_{\; b}$ acting in the super-space
$\cal{V_{(N|M)}}$, namely, the commutation relations
(\ref{CommG}) turn into the commutation relations (\ref{osp06a})
after the change of the definition of the supermetric $\varepsilon_{a b} \to
\varepsilon_{ba} = \epsilon (-1)^{[a]} \varepsilon_{a b}$ (see also the
discussion in Remark 5 below).

\noindent
 {\bf Remark 4.} The conditions (\ref{SymCond})
 for the generators of $osp$ read
in component form  (cf. (\ref{OspAlg})):
\begin{equation} \label{SymmCond1}
G_{ab} +\epsilon(-1)^{[a][b]+[a]+[b]} G_{ba}=0 \, , \;\;\;\;\;
 G_{ab} \equiv \varepsilon_{ac} \, G^c_{\; b} \, .
\end{equation}
In particular, it follows from (\ref{SymCond}), (\ref{SymmCond1})
 that the matrix $G$ is traceless
 $$
 0 = \mathcal{K}_{12} \Bigl(G_1  +
 (-)^{12} G_2 (-)^{12}   \Bigr) \mathcal{K}_{12} =
 2 (\varepsilon_{ab} \, G^a_{\;\, c} \, \bar{\varepsilon}^{cb})
 \, \mathcal{K}_{12} =
 2 \, \epsilon \, {\rm str}(G) \, \mathcal{K}_{12} \; .
 $$
{\bf Remark 5.}
The characteristic identity (\ref{Cond1}) is equivalent to the equation
$$
\mathcal{K}_{12} \Bigl(\beta G_1  + G_1 (-)^{12} G_2 (-)^{12} \Bigr)
= \Bigl( \beta G_1  + (-)^{12} G_2 (-)^{12} G_1   \Bigr) \mathcal{K}_{12} \;
,
$$
provided that the relations (\ref{SymCond}) and (\ref{CommG}) are satisfied.

 \noindent
{\bf Remark 6.} Comparing the $RLL$-relations (\ref{eq:RLL}) and the graded
Yang-Baxter equation (\ref{YBEgr}), one finds that
the latter can be written in the form of the $RLL$-relation with
the $L$-operator represented as an operator
 in ${\cal V}_{(N|M)}^{\otimes 2}$
 \begin{equation} \label{Lfunda}
 L(u) = \frac{1}{u^2} (-)^{12} R_{12}(u) (-)^{12} =
{\bf 1} + \frac{1}{u} \Bigl({\bf 1} \beta +
(\tilde{\cal K} - \epsilon {\cal P}) \Bigr)
 - \frac{\epsilon \beta}{u^2} {\cal P} \; .
\end{equation}
Then the operators $L_1(u)$ and $L_2(v)$ in (\ref{eq:RLL}) should be
understood
 as $\frac{1}{u^2} (-)^{13} R_{13}(u) (-)^{13}$ and
  $\frac{1}{u^2} (-)^{23} R_{23}(u) (-)^{23}$, respectively.
 Taking into account the term proportional to $u^{-1}$
 in (\ref{Lfunda}), we represent
 the traceless generators $G^{a_1}_{\;\; c_1}$ which
 satisfy (\ref{CommG}) as
 \begin{equation} \label{TG}
 G^{a_1 a_2}_{\; c_1 c_2} \equiv
  T^{a_2}_{\; c_2}(G^{a_1}_{\; c_1})  =
  (\tilde{\cal K}^{a_1 a_2}_{\;\; c_1 c_2} -
  \epsilon {\cal P}^{a_1 a_2}_{\;\; c_1 c_2}) =
 \bar{\varepsilon}^{a_2 a_1} \varepsilon_{c_2 c_1} -
  \epsilon (-1)^{[a_1][a_2]} \delta^{a_1}_{\; c_2}
  \delta^{a_2}_{\; c_1} \; ,
 \end{equation}
i.e., this formula gives the defining representation $T$ of the generators
 $G^a_{\; b}$ of $osp$ with the
 structure relations (\ref{CommG}) and the conditions (\ref{SymCond}),
 (\ref{SymmCond1}).  We note that the choice of the basis of $osp$ in
 (\ref{TG})
 differs from the choice of the basis of $osp$ in (\ref{osp03}) by sign
 factors
 $$
 T^{a_2}_{\; c_2}(G^{a_1}_{\; c_1})  =
 (-1)^{[a_1][a_2] + [c_1][c_2]}
 (\tilde{G}^{a_1}_{\; c_1})^{a_2}_{\; c_2} \; .
 $$
This is consistent with the equality
 $\tilde{G}_{12} = G_{21} = (-)^{12} G_{12} (-)^{12}$,
  where $G_{12}$ and $\tilde{G}_{12}$ are defined in
  (\ref{TG}) and (\ref{osp03}) (compare eqs. (\ref{osp06a}), (\ref{osp08b})
  with (\ref{SymCond})).

\section{Super-oscillator
 representation for linear evaluation of \\ $\mathcal{Y}(osp)$}
 \setcounter{equation}0

In this section we intend to construct an explicit representation of
$\mathcal{Y}(osp)$ in which the generators of $osp \subset \mathcal{Y}
(osp)$ satisfy the quadratic characteristic equation \eqref{Cond1}.
 We  follow the approach of \cite{FIKK} and introduce a generalized
 algebra ${\mathcal A}$ of super-oscillators that consists of both bosonic
 and fermionic oscillators simultaneously.

Consider the super-oscillators $c^a$ $(a=1,2,\dots,N+M)$ as
generators of an associative algebra ${\cal A}$ with the defining relation
\begin{equation}
 \label{suposc}
[c^a,c^b]_\epsilon \equiv c^a c^b + \epsilon (-1)^{[a][b]} c^b c^a
=\bar{\varepsilon}^{ab},
\end{equation}
where the matrix $\bar{\varepsilon}^{ab}$ is defined in \eqref{SupMet1}
and \eqref{SupMet0}. In view of (\ref{commzw}), for $\epsilon = -1$,
 the super-oscillators $c^a$ with $[a]=0\; ({\rm mod}2)$
are bosonic and with $[a]=1 \; ({\rm mod}2)$ are fermionic.
For $\epsilon = +1$ the statistics of the super-oscillators $c^a$
is unusual and we will discuss this in more detail in Remark {\bf 8.}
at the end of this section. Nevertheless, we assume the grading
 to be standard grad$(c^a)=[a]$ in both cases $\epsilon = \pm 1$
 and therefore the defining relations (\ref{suposc}) are invariant under
the action $c^a\rightarrow c'^a=U^a_{\ c} c^c$ of the super-group
$OSp$ with the elements $U \in Osp$ (see \cite{FIKK}).

With the help of convention (\ref{agree}) for lowering indices  one can
write relations (\ref{suposc}) in the equivalent forms
\begin{equation}
 \label{suposc1}
[c_a,c_b]_\epsilon \equiv c_a c_b +\epsilon (-1)^{[a][b]} c_b c_a
=\varepsilon_{ba}
\;\; \Leftrightarrow  \;\;
c_a c^b + \epsilon(-1)^{[a][b]} c^b c_a =\delta^b_a \; .
\end{equation}
 The super-oscillators $c^a$ satisfy the following contraction
 identities:
 $$
 c^a c_a = \bar{\varepsilon}^{ab} \varepsilon_{ad}c_b c^d
= \epsilon(-1)^{[a]} c_a c^a \; , \;\;\;
 c_a c^a = \bar{\varepsilon}^{ab} \varepsilon_{ad}c^d c_b
= \epsilon(-1)^{[a]} c^a c_a \; .
$$
 So, we have
\begin{equation} \label{Contract}
 \begin{array}{c}
c^ac_a =
 \frac{1}{2} \bar{\varepsilon}^{ab}
 (c_b c_a + \epsilon(-1)^{[a]} c_a c_b) =
 \frac{1}{2} \bar{\varepsilon}^{ab}
 \varepsilon_{ab} =\frac{\omega}{2} \; , \\ [0.3cm]
 c_a c^a =
 \frac{1}{2} \bar{\varepsilon}^{ab}
 (c_a c_b  + \epsilon(-1)^{[a]} c_b c_a ) =
 \frac{1}{2} \bar{\varepsilon}^{ab}
 \varepsilon_{ba} =\frac{D}{2} \; ,
 \;\;\;\; D \equiv N+M \; .
\end{array}
 \end{equation}
Further we need the super-symmetrised product of two super-oscillators:
\begin{equation}
\label{symcacb}
c^{(a}c^{b)} := \frac{1}{2} \bigl( c^a c^b -\epsilon (-1)^{[a][b]} c^b c^a
\bigr) =  -\epsilon (-1)^{[a][b]} c^{(b}c^{a)}
 \; \in \; {\cal A} \; ,
\end{equation}
and define the operators
\begin{equation}
\label{defF}
F^{ab} \equiv \epsilon (-1)^{[b]} c^{(a}c^{b)}
 \; , \;\;\;\;\  F^{a}_{\ b}=\varepsilon_{bc}F^{ac} \; .
\end{equation}

In \cite{FIKK} we have proved the following statement.
 \begin{proposition}\label{propF}
The operators $F^{ab} \in {\cal A}$ defined in (\ref{defF})
are traceless and possess the symmetry
property (\ref{SymCond}), (\ref{SymmCond1}):
\begin{equation} \label{t1}
\str(F)=(-1)^{[a]}F^{a}_{\ a}=0 \; , \;\;\;\;
F^{ab} = -\epsilon (-1)^{[a][b]+[a]+[b]} F^{ba}.
\end{equation}
In addition
they satisfy the supercommutation relations (\ref{CommG}) for
the generators of
$osp$
\begin{equation} \label{ti2}
(-)^{12} F_1 (-)^{12} F_2  -  F_2 (-)^{12} F_1 (-)^{12}  =
[\epsilon\mathcal{P}_{12}-\tilde{\mathcal{K}}_{12}, \; F_2 ] \; ,
\end{equation}
and obey the quadratic characteristic identity
(\ref{Cond1}):
\begin{equation} \label{t3}
F^a_{\ b}F^{b}_{\ c}+\beta F^a_{\ c} -\frac{\epsilon}{\omega} \str(F^2)
\delta^a_c
=0,
\end{equation}
where $\beta = 1 - \omega/2$.
\end{proposition}
 Thus, the elements $F^{a}_{\;\; b} =\epsilon \, \varepsilon_{bd} (-1)^{[b]}
  c^{(a} c^{d)} \in {\cal A}$ given in (\ref{defF}) form
 a set of traceless generators
 of $osp$ which satisfy all conditions of Proposition
 {\bf \ref{prop2a}} and it means  that the following statement
 holds.
   \begin{proposition}\label{propB}
  The $L$-operator (\ref{L01}) in the super-oscillator
   representation (\ref{suposc}):
 \begin{equation} \label{eq:Losc}
L^a_{\; b}(u) =(u+\alpha-\frac12)\delta^a_b+\epsilon(-1)^{[b]}
c^ac_b \equiv(u+\alpha-\frac12)\delta^a_b+B^a_{\;b} \; ,
\end{equation}
where we introduce for convenience
$B^a_{\;b} \equiv F^a_{\;b} + \frac12\delta^a_b = \epsilon(-1)^{[b]}
c^ac_b$, obey the RLL equation (\ref{eq:RLL}) which in the component
form is given by
\begin{equation} \label{eq:RLLc}
(-1)^{[c_1]([b_2]+[c_2])}R^{a_1a_2}_{b_1b_2}(u-v)L^{b_1}_{c_1}
(u)L^{b_2}_{c_2}(v)=(-1)^{[a_1]([a_2]+[b_2])} L^{a_2}_{b_2}(v)
L^{a_1}_{b_1}(u)R^{b_1b_2}_{c_1c_2}(u-v) \; ,
\end{equation}
and the $R$-matrix (\ref{Rmatri}) is
$$
R^{a_1a_2}_{b_1b_2}(u)=u(u+\beta)\delta^{a_1}_{b_
1}\delta^{a_2}_{b_2}-\epsilon(u+\beta)(-1)^{[a_1][a_2]}\delta
^{a_1}_{b_2}\delta^{a_2}_{b_1}+u\bar\varepsilon^{a_1a_2}
\varepsilon_{b_1b_2} \; .
$$
\end{proposition}
\noindent
{\bf Proof.} One can prove this Proposition directly.
To simplify the notation, we  write $(-1)^a$ and $(-1)^{ab}$
instead of $(-1)^{[a]}$ and $(-1)^{[a][b]}$.
After substituting the $L$-operator (\ref{eq:Losc}),
 the RLL equation (\ref{eq:RLLc}) takes the form
$$
(u-v)(u-v+\beta)\Big((-1)^{c_1(a_2+c_2)}B^{a_1}_{\;c_1}
B^{a_2}_{\;c_2}-(-1)^{a_1(a_2+c_2)}B^{a_2}_{\;c_2}
B^{a_1}_{\;c_1}\Big)-
$$
$$
-(u-v+\beta)\epsilon(-1)^{a_1a_2+a_1c_1+c_1c_2}\Big(u\delta^
{a_2}_{c_1}B^{a_1}_{\;c_2}+v\delta^{a_1}_{c_2}B^{a_2}_{\;c_1}-
u\delta^{a_1}_{c_2}B^{a_2}_{\;c_1}-v\delta^{a_2}_{
c_1}B^{a_1}_{\;c_2}\Big)+
$$
$$
+(u-v)\bar\varepsilon^{a_1a_2}\Big((-1)^{c_1(c_2+b_2)}
\varepsilon_{b_1b_2}(u\delta^{b_1}_{c_1}+B^{b_1}_{\;c_1})B^{
b_2}_{\;c_2}+v\varepsilon_{b_1c_2}B^{b_1}_{\;c_1}\Big)-
$$
$$
-(u-v)\varepsilon_{c_1c_2}\Big((-1)^{a_1(a_2+b_2)}\bar
\varepsilon^{b_1b_2}B^{a_2}_{\;b_2}(u\delta^{b_1}_{c_1}+ B^{
b_1}_{\;c_1})+v\bar\varepsilon^{b_1a_2}B^{a_1}_{\;b_1}\Big)=0.
$$
Taking into account the representation
 $B^a_{\;b}=\epsilon(-1)^bc^ac_b$ and
defining relations (\ref{suposc}), one checks that the above
relation is valid at arbitrary $u$ and $v$. \hfill \qed

  \noindent
{\bf Remark 7.} The Quadratic Casimir operator $C_2$
of the superalgebra $osp(N|M)$ in the differential
representation (\ref{defF}) is equal to the fixed number
\begin{equation}
 \label{kasi}
 C_2 = (-1)^{[a]} F^a_{\;\; b} F^b_{\;\; a} =
\frac{\epsilon}{4}\omega (\omega -1) \; .
\end{equation}
It means that this realization (\ref{defF}) corresponds to a limited class of
representations of the superalgebra $osp(N|M)$. This fact reflects the general
statement of \cite{Drinf} that not all representations of
simple Lie algebras $\mathfrak{g}$ of $B,C$ and $D$ types
 are the representations of the corresponding Yangians $Y(\mathfrak{g})$.

\vspace{0.2cm}

\noindent
{\bf Remark 8.}
 For $\epsilon=-1$ and even $M=2m$ the super-oscillator algebra (\ref{suposc})
is represented in terms
of $m$ copies of the bosonic Heisenberg algebras $c^{j}=x^j$,
$c^{m+j}=\partial^j$, $j=1,\ldots,m$, and $N$ fermionic
oscillators $c^{2m +\alpha} = b^\alpha$, $\alpha=1,2,\ldots,N$,
with the (anti)commutation relations
 \begin{equation} \label{eq:ib4}
 [x^i, \; \partial^j] = -\delta^{ij} \; ,\;\;\;\; [b^\alpha,b^\beta]_+ :=
b^\alpha\,b^\beta + b^\beta\,b^\alpha = 2 \delta^{\alpha\beta}
\; ,\;\;\;\;
 [x^i, \; b^\alpha]= 0 = [\partial^i, \; b^\alpha] \; ,
\end{equation}
 which are equivalent to (\ref{suposc}) with the
 choice of the metric $\bar{\varepsilon}^{ab}$
 as $(M+N)\times (M+N)$ matrix
\begin{equation} \label{eq:eps5}
 \bar{\varepsilon}^{ab} = {\footnotesize  \left(\!\!
 \begin{array}{ccc}
 0 & -I_m & 0 \\
  I_m & 0 & 0 \\
   0 & 0 & 2 \, I_N
 \end{array} \!\! \right)} \;\;\; \Rightarrow \;\;\;
 \varepsilon_{ab} = {\footnotesize  \left(\!\!
 \begin{array}{ccc}
 0 & I_m & 0 \\
  -I_m & 0 & 0 \\
   0 & 0 & \frac{1}{2}I_N
 \end{array} \!\! \right)} \; .
\end{equation}
The fermionic variables $b^\beta$ with the commutation relations
(\ref{eq:ib4}) generate the $N$-dimensional Clifford algebra.
Let $N$ be an even number $N=2n$. In this case, one can introduce
the longest element $b^{(N+1)}= (i)^n \, b^1 b^2 \cdots b^{N}$ which anticommutes
with all generators $b^\alpha$ and possesses $(b^{(N+1)})^2=1$.
Then, for $\epsilon=+1$ and even numbers $M=2m$, $N=2n$,
 one can realize the super-oscillator algebra (\ref{suposc})
 (with the metric (\ref{eq:eps5})) in terms of the generators
 \begin{equation} \label{eq:ib41}
c^{j}=x^j \cdot b^{(N+1)} \; , \;\;\;
c^{m+j}=\partial^j \cdot b^{(N+1)} \;\;\;(j=1,\ldots,m)
\; , \;\;\;
c^{2m +\alpha} = b^\alpha \;\;\; (\alpha=1,2,\ldots,N) \; ,
 \end{equation}
 where the operators $x^i$, $\partial^j$ and
 $b^\alpha$ satisfy (\ref{eq:ib4}).
Note that the super-oscillator algebra (\ref{suposc}) for $\epsilon=+1$
has an unusual property that generators $c^a$ and $c^b$
with gradings $[a]=0$ and $[b]=1$ anticommute, which is not usual
feature of bosons and fermions in field theories.

The implementation (\ref{eq:ib4}) of algebra (\ref{suposc})
 suggests the rules of Hermitian conjugation for the generators $c^a$
  \begin{equation} \label{herm}
  \begin{array}{c}
 (c^{j})^\dagger = c^{j} \, , \;\;\;\;
 (c^{m+j})^\dagger  = - c^{m+j} \, , \;\;\;\;
 j=1,\ldots,m \; , \\ [0.2cm]
  (c^{2m +\alpha})^\dagger  = c^{2m +\alpha}
  \; , \;\;\; \alpha=1,2,\ldots,N \; ,
 \end{array}
\end{equation}
which follow from the commonly used properties of the Heisenberg and
Clifford algebras: $(x^j)^\dagger =x^j$,
$(\partial^j)^\dagger = - \partial^j$,
$(b^\alpha)^\dagger = b^\alpha$, $(b^{(N+1)})^\dagger = b^{(N+1)}$.
We shall apply the rules (\ref{herm}) below.

\vspace{0.2cm}

\noindent
{\bf Remark 9.} Consider the graded tensor product ${\cal A} \otimes
{\cal A}$ and denote the generators of the first and second factors
 in ${\cal A} \otimes {\cal A}$ respectively as $c_1^a$ and $c_2^a$.
 Since $\otimes$ is the graded tensor product, we have (cf. (\ref{suposc}))
 \begin{equation}
 \label{eq:c4}
 [c_1^a , \; c_2^b]_\epsilon \equiv
  c_1^ac_2^b + \epsilon(-1)^{ab}c_2^bc_1^a = 0 \; .
 \end{equation}
 Any element of
 ${\cal A} \otimes {\cal A}$ can be written as a
 polynomial $f(c^a_1,c^b_2)$ and its
 condition of invariance under the action of the group $Osp$
 is written as
 $$
 \left[A_{ba}(F_1^{ab} + F_2^{ab}),
 \; f(c^a_1,c^b_2) \right]  = 0,
 $$
 where (see (\ref{defF}))
 \begin{equation}
\label{defF12}
F_1^{ab} \equiv \epsilon (-1)^{b}c_1^{(a}c_1^{b)}
 \; , \;\;\;\  F_2^{ab} \equiv \epsilon (-1)^{b}c_2^{(a}c_2^{b)} \;
\end{equation}
are the generators of the $osp$ algebras and $A_{ba}$ are the super-
parameters (with  ${\rm grad}(A_{ba})= [a]+[b]$).
 In the case of an even function
 $f$, when ${\rm grad}(f)=0$, this invariance condition is
 equivalent to
 \begin{equation}
\label{AFc3}
\left[ (F_1^{ab} + F_2^{ab}), \; f(c^a_1,c^b_2) \right]  = 0 \; .
\end{equation}

 \vspace{0.3cm}

Now we introduce the super-symmetrized product
$c^{(a_1}\cdots c^{ a_k)}$ of any
number of super-oscillators, which generalizes the
super-symmetrized product of two super-oscillators (\ref{symcacb}).
The general definition and properties of such super-symmetrized products
 are given in Appendix {\bf \ref{symprod}}.
 In  \cite{FIKK} we have proved the following statement.
\begin{proposition} \label{propFc}
The elements
 \begin{equation}
 \label{invars}
 I_k = \varepsilon_{a_1b_1}\dots \varepsilon_{a_kb_k} \;
 c_1^{(a_1}\cdots c_1^{ a_k)} c_2^{(b_k} \cdots c_2^{b_1)}
 \; \in \; {\cal A}\otimes {\cal A} \; , \;\;\;\; k=1,2,\dots \; ,
  \end{equation}
 are invariant under the action (\ref{transU})
 of the supergroup $OSp$:
 $c^a \;\; \to \;\; U^a_{\;\; b} \, c^b$.
 It means that the elements (\ref{invars}) are
 invariant under the action of the Lie superalgebra  $osp$ and
 satisfy the invariance condition (\ref{AFc3}):
\begin{equation}
 \label{invarC}
\left[\varepsilon_{a_1b_1}\dots
\varepsilon_{a_kb_k} c_1^{(a_1}\cdots c_1^{a_k)}
c_2^{(b_k} \cdots c_2^{b_1)} ,F_1^{ab} +F_2^{ab}\right] = 0 \; ,
\end{equation}
 where $F_1^{ab}$ and $F_2^{ab}$ are the generators (\ref{defF12}) of
 the Lie super-algebra $osp$  (see Proposition {\bf \ref{propF}}).
\end{proposition}
It turns out that the invariants (\ref{invars})
are not functionally independent. Indeed, we have
the following statement.
\begin{proposition} \label{reqinv}
The invariants (\ref{invars}) satisfy the recurrence relation
\begin{equation} \label{eq:zim1}
I_k \, I_1 = I_{k+1}+\frac k4\big((k-1) - \omega\big) I_{k-1}
\, ,\qquad
 \o=\epsilon (N-M) \; ,
\end{equation}
where $I_0=1$ and $I_1= \varepsilon_{ab} c_1^a \, c_2^b
 = c_1^a \, c_{2a}$. In the representations
 (\ref{eq:ib4}), (\ref{eq:ib41}) and (\ref{eq:eps5})
 Hermitian conjugations of invariant elements (\ref{invars}) are
 \begin{equation} \label{herm1}
 I_{2k}^\dagger = I_{2k} \, ,\qquad  I_{2k+1}^\dagger = - I_{2k+1} \; .
\end{equation}
\end{proposition}
\noindent
{\bf Proof.} The derivation of the recurrence relation
(\ref{eq:zim1}) is given in Appendix {\bf \ref{symprod}}.
 To prove (\ref{herm1}), it is useful to define the invariants
 \begin{equation}\label{invtil}
 \tilde{I}_m = \sigma^m \, I_m \; ,
 \end{equation}
 where $\sigma^2=-1$, i.e. $\sigma=\pm i$.
Then,
the recurrence relation (\ref{eq:zim1}) for new invariants
$\tilde{I}_k$ has the form:
\begin{equation}\label{rri}
\tilde{I}_{k+1}=z \tilde{I}_k+
 \frac k4(k-1-\o)\tilde{I} _{k-1},\qquad\qquad
 \o=\epsilon (N-M) \; ,
\end{equation}
 where $\tilde{I}_0=1$ and we introduce the operator
 \begin{equation}\label{invz}
z:= \tilde{I}_1 = \sigma I_1
=\sigma\e_{ab}c_1^ac_2^b=\sigma(c_1\cdot c_2)=-\sigma
\e_{ba}c_2^bc_1^a=-\sigma  (c_2\cdot c_1),
\end{equation}
which is Hermitian $z^\dagger = z$ in the representations
(\ref{eq:ib4}), (\ref{eq:ib41}) and (\ref{eq:eps5}).
 One can prove the latter statement by making use of the rules
 (\ref{herm}) and commutation relations (\ref{eq:c4}). In view
 of the recurrence relation (\ref{rri}) and initial conditions
 $\tilde{I}_0=1$ and $\tilde{I}_1= z$ all invariant operators
 $\tilde{I}_k$ are $k$-th order polynomials (with real coefficients)
 of the Hermitian operator $z$. Therefore all $\tilde{I}_k$
 are the Hermitian operators $\tilde{I}_k^\dagger = \tilde{I}_k$,
 and therefore, taking into account (\ref{invtil})
 and $\sigma^*=-\sigma$, we deduce (\ref{herm1}). \hfill \qed

 \vspace{0.2cm}

 Now we introduce a generating function of the Hermitian
 invariant operators $\tilde{I}_k$:
\begin{equation}\label{rf}
F(x|z)=\sum_{k=0}^\infty\tilde{I}_k\frac{x^k}{k!} \, .
\end{equation}
Since the invariants $\tilde{I}_k$ are polynomials in $z$,
the generating function (\ref{rf}) depends on $x$ and $z$ only.
\begin{proposition} \label{genfun}
The generating function (\ref{rf}) is equal to
\be\lb{rf3}
F(x|z)= \Bigl(1-\frac x2\Bigr)^{\frac\o2-z}
\Bigl(1+\frac x2\Bigr)^{\frac\o2+z} \; .
\ee
\end{proposition}
\noindent
{\bf Proof.} Using the recurrence relation (\ref{rri})
 we obtain:
\be\lb{rf1}
\sum_{k=0}^\infty\tilde{I}_{k+1}\frac{x^k}{k!}=z\sum_{k=0}^\infty \tilde{I}
_k\frac{x^k}{k!}+\frac14\sum_{k=2}^\infty\tilde{I}_{k-1}\frac{x^k}{(k-2)!}-
\frac\o4\sum_{k=1}^\infty\tilde{I}_{k-1}\frac{x^k}{(k-1)!}.
\ee
Now changing the summation indices and using (\ref{rf}) one deduces:
\be\lb{rf2}
F_x(x|z)=zF(x|z)+\frac{x^2}4F_x(x|z)-\frac{x\o}4F(x|z) \, ,
\ee
where $F_x(x
|z)\equiv\dd_xF(x|z)=\sum_{k=0}^\infty\tilde{I}_{k+1}\frac{x^k}{k!}$.
The general solution to this ordinary differential equation
 is given in (\ref{rf3}) up to an arbitrary constant
 factor $c$.
 The invariants $\tilde{I}_k$
 are extracted from the generating function (\ref{rf})
 using the formula
\be\lb{ikk}
\tilde{I}_k(z)=\dd^k_xF(0|z)=
c \; \dd^k_x \, (1-\frac x2)^{\frac\o2-z}(1+\frac x2)^{\frac\o2+z}\Big|_{x=0},
\ee
from which we fix the constant $c=F(0|z)=\tilde{I}_0=1$. \hfill \qed

\section{The construction of the R-operator in the super-oscillator
representation\label{RFRF}}
\setcounter{equation}0

 Let $T$ be the defining representation of the Yangian $Y(osp)$.
In the previous section we have considered the $RLL$-relation (\ref{eq:RLL})
and (\ref{eq:RLLc}) that intertwines $L$-operators $||L^{a}_{\;\; b}(u)||
\in T(Y(osp)) \otimes {\cal A}$ (given in (\ref{eq:Losc}))
by means of the $R$-matrix (\ref{Rmatri}) in the defining representation, i.e.,
$R(u) \in T(Y(osp)) \otimes T(Y(osp))$. In other words, the  $R$-matrix
in the $RLL$-relations (\ref{eq:RLL}) and (\ref{eq:RLLc}) acts
in the space ${\cal V}_{(N|M)}^{\otimes 2}$, where ${\cal V}_{(N|M)}$
is the space of the defining representation $T$ of $Y(osp(N|M))$.

There is another type of $RLL$-relations which intertwines the
$L$-operators (\ref{eq:Losc}) by means of the $R$-matrix
in the super-oscillator representation, i.e., $\hat{\mathcal{R}}(u)\in{\cal A}
\otimes {\cal A}$, where $\otimes$ is the graded tensor product.
In components, this type of $RLL$ relations has the form
\begin{equation} \label{eq:RLL2}
\hat{\mathcal{R}}_{12}(u) {L_1}^a_{\;b}(u+v){L_2}^b_{\;c}(v) =
{L_1}^a_{\;b}(v){L_2}^b_{\;c}(u+v)\hat{\mathcal{R}}_{12}(u) ,
\end{equation}
or after substitution of the $L$-operator (\ref{eq:Losc})
we have
 \begin{equation}
\label{RLL51}
\begin{array}{c}
\hat{\mathcal{R}}_{12}(u)\big((u+v)\delta^a_b+\epsilon(-1)^bc_1^a
{c_1}_b\big)\big(v\delta^b_c+\epsilon(-1)^cc_2^b{c_2}_c\big)= \\ [0.2cm]
=\big(v\delta^a_b+\epsilon(-1)^bc_1^a{c_1}_b\big)\big((u+v)\delta
^b_c+\epsilon(-1)^cc_2^b{c_2}_c\big)\hat{\mathcal{R}}_{12}(u) .
\end{array}
\end{equation}
Here for simplicity we fix $\alpha = 1/2$ in the definition of the
 $L$-operators and associate the first and second factors in
${\cal A} \otimes {\cal A}$, respectively, with the algebras
 ${\cal A}_1$ and ${\cal A}_2$ generated
 by the elements $c_1^a$ and $c_2^b$ such that
 $[c_1^a, \; c_2^b]_\epsilon = 0$ (see (\ref{eq:c4})).

The $RLL$ relation (\ref{RLL51}) is quadratic with respect
to the parameter $v$. The terms proportional to $v^2$ are cancelled,
the terms proportional to $v$ give
\begin{equation} \label{eq:RLL3}
\hat{\mathcal{R}}_{12}(u)(c_1^a{c_1}_c+c_2^a{c_2}_c)
=(c_1^a{c_1}_c+c_2^a{c_2}_c)\hat{\mathcal{R}}_{12}(u),
\end{equation}
while the terms independent of $v$ are
\begin{equation} \label{eq:RLL4}
\hat{\mathcal{R}}_{12}(u)\big(u\delta^a_b+\epsilon(-1)^bc_1^a
{c_1}_b\big)(-1)^cc_2^b{c_2}_c= (-1)^bc_1^a{c_1}_b\big(u\delta
^b_c+\epsilon(-1)^cc_2^b{c_2}_c\big)\hat{\mathcal{R}}_{12}(u) .
\end{equation}

\subsection{The Shankar-Witten form of the $R$ operator}

The relations (\ref{eq:RLL3}) are nothing but the invariance conditions
(\ref{AFc3})  with respect to the adjoint action of $osp$
\begin{equation} \label{eq:Rcond1}
\left[ \hat{\mathcal{R}}(u), F_1^{ab} +F_2^{ab} \right] = 0 \; .
\end{equation}
It means that one can search for the $\hat{\mathcal{R}}(u)$-operator
as a sum of $osp$-invariants (\ref{invars})
\begin{equation}
 \label{anzR}
\hat{\mathcal{R}}_{12}(u) = \sum_k \frac{r_k(u)}{k!} \, I_k
= \sum_k \frac{r_k(u)}{k!} \,
\varepsilon_{\vec{a},\vec{b}} \
 c_1^{(a_1\dots a_k)} c_2^{(b_k\dots b_1)} \; ,
\end{equation}
where we use the concise notation
$$
\varepsilon_{\vec{a} ,\vec{b}} = \varepsilon_{a_1b_1}\dots
\varepsilon_{a_kb_k}  ,
\;\;\;\; c_1^{(a_1\dots a_k)} := c_1^{(a_1} \cdots c_1^{a_k)}, \;\;\;\;
c_2^{(b_k\dots b_1)} := c_2^{(b_k}\cdots c_2^{b_1)}  .
$$
Inserting this ansatz into the condition
 \eqref{eq:RLL4}, we obtain (see \cite{FIKK})
 the recurrence relation for $r_k(u)$
\begin{equation} \label{Rrec1}
r_{k+2}(u) = \frac{4(u-k)}{k+2+u-\omega} r_k(u) \; ,
\end{equation}
which is solved in terms of the $\Gamma$-functions:
\begin{equation}
 \label{sol-rm}
 \begin{array}{rl}
r_{2m}(u) &= (-4)^{m}
\frac{\Gamma(m-\frac{u}{2})}{\Gamma(m+1+\frac{u-\omega}{2})}
A(u), \\
r_{2m+1}(u) &= (-4)^{m}
\frac{\Gamma(m-\frac{u-1}{2})}{\Gamma(m+1+\frac{u-\omega+1}{2})} B(u)  \; ,
\end{array}
\end{equation}
where the parameter $\omega = \epsilon(N-M)$ was defined
in (\ref{oMN}) and $A(u),B(u)$ are arbitrary functions of $u$.
Substituting (\ref{sol-rm}) in (\ref{anzR}) gives
the expression for the $osp$-invariant $R$-matrix which
intertwines two $L$ operators in (\ref{eq:RLL2}).


The methods used in \cite{FIKK} (for derivation of (\ref{anzR})
and  (\ref{sol-rm})) require the introduction
of additional auxiliary variables and are technically quite nontrivial
 and cumbersome. Below in this paper, in Section {\bf \ref{relapp}},
 we give a simpler and more elegant derivation of conditions
 (\ref{Rrec1}). This derivation is based on an application of the generating
 function (\ref{rf3}) for the invariants $\tilde{I}_k$, where
 the explicit form (\ref{rf3}) is obtained
  by means of the recurrence relation (\ref{eq:zim1}).

\subsection{The Faddeev-Takhtajan-Tarasov type $R$ operator}


There is another form of $R$ operators which intertwines the
$L$ operators in the $RLL$ equations (\ref{RLL51})
and are expressed as a ratio of Euler Gamma-functions.
For the $s\ell(2)$ case this type of solutions for $R$ operator
was first obtained in \cite{FTT} (see also \cite{Fadd} and \cite{Der1}).
The generalization to the $s\ell(N)$ case (for a wide
class of representations of $s\ell(N)$) was given in \cite{DerMan}.
For orthogonal and symplectic algebras (and a very special class of their
representations) analogous solutions of  (\ref{RLL51}) were
recently obtained in \cite{KaKi}. Below we generalize the results of
\cite{KaKi} and find the solutions for the super-oscillator
Faddeev-Takhtajan-Tarasov type $R$-operator in the case of
$osp$ Lie superalgebras.


\begin{proposition}
The $R$ operator intertwining the super-oscillator $L$ operators in the
$RLL$ equations (\ref{eq:RLL2}), (\ref{RLL51})
 obeys the finite-difference equation
\begin{equation} \label{eq:R7}
 \hat{\mathcal{R}}_{12}(u|z+1)\; (z - u) =
\hat{ \mathcal {R}}_{12}(u|z-1) \; (z +u) \; ,
 \end{equation}
 where $z = \sigma \, c_1^a \, c_{2a}$ and $\sigma^2=-1$.
The solution of this
 functional equation is given by the ratio of the
 Euler Gamma-functions
\begin{equation} \label{eq:R6}
\hat{\mathcal{R}}_{12}(u|z)=r(u,z)\frac{\Gamma\big(\frac12(z+1+u)
\big)}{\Gamma\big(\frac12(z+1-u)\big)} ,
\end{equation}
 where $r(u,z)$ is an arbitrary periodic function
 $r(u,z+2)=r(u,z)$ which normalizes the solution.
\end{proposition}

\noindent
{\bf Proof.}
Taking into account the experience related to the orthogonal and
symplectic cases (see \cite{KaKi}), we will look for a
solution to the first equation (\ref{eq:RLL3}) as
\begin{equation} \label{eq:Rh2}
\hat{\mathcal{R}}_{12}(u)=\hat{\mathcal{R}}_{12}(u|z) ,\qquad
z=\sigma c_1^a{c_2}_a=\epsilon\sigma (-1)^a c_{1a}c_2^a
= - \sigma c_2^a c_{1a} \; ,
\end{equation}
where $\sigma$ is a numerical constant to be defined.
In the last chain of equalities we have used (\ref{eq:c4}). In other
words, the operator $\hat{\mathcal{R}}_{12}(u)$ acting
in ${\cal{V}}_1\otimes{\cal{V}}_2$ is given by a function of an
invariant $z$ bilinear in super-oscillators $c_1^a$ and $c_2^a$.
Note that in the orthogonal and symplectic cases \cite{KaKi} the
conventional invariants $I_k$ (\ref{invars})  are in one-to-one
correspondence with polynomials of $z$ of the order $k$. In the
super-symmetric case of the algebras $osp$ we prove this fact in Appendix
{\bf \ref{symprod}} (see eq. (\ref{eq:zim}) and
comment after this equation). To justify the ansatz (\ref{eq:Rh2}),
we recall that the super-oscillators belonging to different
factors in ${\cal A} \otimes {\cal A}$ and
acting in different auxiliary spaces
${\cal{V}}_1$ and ${\cal{V}}_2$  commute according to (\ref{eq:c4})
\begin{equation} \label{eq:c3}
c_1^ac_2^b= -\epsilon (-1)^{ab}c_2^bc_1^a,
\end{equation}
so we have
\begin{equation} \label{eq:zc1}
\!\! zc_1^b=\sigma c_1^ac_{2a}c_1^b=-\epsilon (-1)^{ab}\sigma
c_1^ac_1^bc_{2a}=(-1)^{ab+1}\sigma
\big(\epsilon \bar\varepsilon^{ab}-(-1)^{ab}c_1^bc_1^a\big)c_{2a} =
c_1^b z-\sigma c_2^b,
\end{equation}
\begin{equation} \label{eq:zc2}
zc_2^b=\sigma c_1^ac_{2a}c_2^b=c^a_1\sigma
(\delta^b_a-\epsilon(-1)^{ab}c_2^bc_2^a)=\sigma c_1^b-\sigma\epsilon
(-1)^{ab}c_1^ac_2^bc_{2a}=\sigma c_1^b+ c_2^bz.
\end{equation}
Combining these relations we obtain
\begin{equation} \label{eq:zc12}
z\, (c_1^ac_{1b}+c_2^ac_{2b})=
 (c_1^ac_{1b}+c_2^ac_{2b})\, z,
\end{equation}
i.e. $z$ commutes with the sum $c_1^ac_{1b}+c_2^ac_{2b}$,
and hence an arbitrary function $\hat{\mathcal{R}}_{12}(u|z)$
depending on $z$ satisfies the invariance
conditions (\ref{eq:RLL3}) and (\ref{eq:Rcond1}).

Let us introduce
\begin{equation} \label{cpm5}
c^b_\pm := (c_1^b \pm \sigma c_2^b) \; ,
\end{equation}
 and consider a linear combination of (\ref{eq:zc1}) and
(\ref{eq:zc2})
\begin{equation} \label{eq:cpm}
zc^b_\pm\equiv z(c_1^b\pm\sigma c_2^b)= c^b_\pm z
\pm \sigma^2(c_1^b\mp \sigma^{-1} c^b_2)= c^b_\pm(z\mp 1),
\end{equation}
where the last equation is obtained under the choice
\begin{equation} \label{eq:se}
\sigma^2 = -1 \;\;\;\; \Rightarrow \;\;\;\;
\sigma=\sqrt{-1}=\left\{\ba{cc}i \; , \\
-i \; .\ea\right.
\end{equation}
Taking into account  (\ref{eq:cpm}), we have
\begin{equation} \label{eq:rc}
\hat{\mathcal{R}}_{12}(u|z)\, c^b_\pm=c^b_\pm \,
\hat{\mathcal{R}}_{12}(u|z\mp 1) \; ,
 \qquad c^b_\pm\hat{\mathcal{R}}_{12}(u|z)
 =\hat{\mathcal{R}}_{12}(u|z\pm 1) c^b_\pm \; .
\end{equation}
Then multiplying (\ref{eq:RLL4}) by $c^d_\pm \varepsilon_{da}$
(or by $c^d_\mp \varepsilon_{da}$) from the left and by $c^c_\pm$
 from the right and contracting
oscillator vector indices, one obtains four
independent scalar relations. Two
of them are
\begin{equation} \label{eq:RLL5}
c^d_{\pm}\varepsilon_{da}\hat{\mathcal{R}}_{12}(u|z)
\big(u\delta^a_b+\epsilon(-1)^bc_1^a{c_1}_b\big)(-1)^cc_2^b
{c_2}_c c^c_\pm=
\end{equation}
$$
=c^d_{\pm}\varepsilon_{da}(-1)^bc_1^a{c_1}_b\big( u
\delta ^b_c+\epsilon(-1)^cc_2^b{c_2}_c\big)\hat{\mathcal {R}}
_{12}(u|z) c^c_\pm \; .
$$
Applying (\ref{eq:rc}), (\ref{Contract}),
the definition (\ref{eq:Rh2}) of $z$ and
$$
\begin{array}{c}
c^d_\pm c_{2d} = \sigma(- z \pm \frac{\omega}{2}), \;\;
c^d_\pm c_{1d} = \frac{\omega}{2} \mp z, \;\;
(-1)^c c_{2c} c^c_\pm  =
\epsilon \sigma(z \pm \frac{\omega}{2}),  \;\;
(-1)^c c_{1c} c^c_\pm  =
\epsilon ( \frac{\omega}{2} \pm z),
\end{array}
$$
these two relations (\ref{eq:RLL5}) turn to be functional equations on
 $\hat{\mathcal{R}}_{12}(u|z)$:
$$
\begin{array}{c}
\hat{\mathcal{R}}_{12}(u|z\pm1)c^d_\pm
\varepsilon_{da}\big(u\delta^a_b
+\epsilon(-1)^bc_1^a{c_1}_b\big)
(-1)^cc_2^b{c_2}_c c^c_\pm = \\ [0.2cm]
=c^d_\pm \varepsilon_{da}(-1)^bc_1^a{c_
1}_b\big( u\delta ^b_c+\epsilon(-1)^cc_2^b{c_2}_c\big)c^c_\pm \hat{ \mathcal {R}}_{12}(u|z\mp1) \; ,
\end{array}
\quad\Rightarrow
$$
$$
\hat{\mathcal{R}}_{12}(u|z\pm1)\epsilon(u \mp z)
\big(z^2 - \frac{\omega^2}{4}\big) =
\epsilon(-u\mp z)\big(z^2 - \frac{\omega^2}{4}\big)
\hat{ \mathcal {R}}_{12}(u|z\mp 1) \; .
$$
 Canceling the common factor
 $\epsilon \big(z^2 - \frac{\omega^2}{4}\big)$ in both sides
  we obtain a pair of equations
 \be
 \lb{eq:R8}
   \hat{\mathcal{R}}_{12}(u|z\pm1)\; (u \mp z) =
(-u\mp z)\; \hat{ \mathcal {R}}_{12}(u|z\mp 1) \; ,
 \ee
  which are equivalent for both choices of signs
 to the one equation (\ref{eq:R7}).
In a similar fashion the other pair of relations gives identities
 \begin{equation} \label{eq:RLL6}\begin{array}{c}
 c^d_{\mp}\varepsilon_{da}\hat{\mathcal{R}}_{12}(u|z)
 \big(u\delta^a_b+\epsilon(-1)^bc_1^a{c_1}_b\big)(-1)^cc_2^b
 {c_2}_cc^c_\pm= \\ [0.3cm]
 =c^d_{\mp}\varepsilon_{da}(-1)^bc_1^a{c_1}_b\big( u
 \delta ^b_c+\epsilon(-1)^cc_2^b{c_2}_c\big)
 \hat{\mathcal {R}}_{12}(u|z)c^c_\pm,
 \end{array}  \quad\Rightarrow
 \end{equation}
$$
\begin{array}{c}
\hat{\mathcal{R}}_{12}(u|z\mp1) c^d_{\mp}\varepsilon_{da}
\big(u\delta^a_b+\epsilon(-1)^bc
_1^a{c_1}_b\big)(-1)^cc_2^b{c_2}_c c^c_\pm =
 \\ [0.3cm]
= c^d_{\mp}\varepsilon_{da} (-1)^bc_1^a{c_
1}_b\big( u\delta ^b_c+\epsilon(-1)^cc_2^b{c_2}_c\big)c^c_\pm
\hat{ \mathcal {R}}_{12}(u|z\mp 1)),
\end{array}
 \quad\Rightarrow
$$
$$
\hat{\mathcal{R}}_{12}(u|z\mp1)\epsilon(u\pm z)
\big(z \pm \frac{\omega}{2} \big)^2=
\epsilon(u\pm z)\big(z \pm \frac{\omega}{2} \big)^2
\hat{ \mathcal {R}}_{12}(u|z\mp1 ),
$$
which are satisfied automatically.
Finally, the solution of the
 functional equations (\ref{eq:R7}), (\ref{eq:R8}) can be found
 immediately and is given in (\ref{eq:R6}) by the ratio of the
 Euler Gamma-functions. \hfill \qed

 \vspace{0.2cm}

We see that the scalar projections
(\ref{eq:RLL5}) and (\ref{eq:RLL6}) of the RLL relation
are exactly the same as in the non-supersymmetric case \cite{KaKi}, i.e.
 no signs related to grading appear.
 Moreover, we stress
  that the functional equation (\ref{eq:R7}) is independent
 of the parameter $\epsilon$, which distinguishes the cases
 of the algebras $osp(N|M)$ and $osp(M|N)$.

 \vspace{0.2cm}
 \noindent
 {\bf Remark 10.} We have two choices (\ref{eq:se}) of the parameter $\sigma$
 and therefore we have two versions of the solution (\ref{eq:R6})
\begin{equation} \label{eq:R61}
\hat{\mathcal{R}}^{(\pm)}_{12}(u|z)=r^{(\pm)}(u,z) \;
\frac{\Gamma\big(\frac12(\pm i c_1^a c_{2a} +1+u)
\big)}{\Gamma\big(\frac12(\pm i c_1^a c_{2a} +1-u)\big)} .
\end{equation}
In view of the identity
 $\Gamma(1-x) \Gamma(x) = \pi/\sin(\pi x)$,
 these two versions are equivalent to each other up to a special choice
 of the normalization functions $r^{(\pm)}(u,z)$. So one can consider
 only one of the solutions (\ref{eq:R61}).

\section{The $ \hat{\mathcal{R}}$ operator  in special cases of
$osp(N|2m)$}
\setcounter{equation}0
In this section, we work out the explicit form of the solution
(\ref{eq:R6}) in a few particular cases.

\subsection{The case of $osp(M|N)=osp(1|2)$}
In this case, we have $N=2$ and $M=1$, and
the superalgebra $osp(1|2)$ is described by the bosonic oscillator
$c^1\equiv a^\dagger,c^2\equiv a$ (in the holomorphic representation
we have $c^1\equiv x,c^2\equiv \partial$) and by one
fermionic variable $c^3 \equiv b$ with the
 commutation relations (\ref{suposc}):
 \begin{equation} \label{eq:ib1}
 [x, \; \partial] = -1 \; ,\qquad \{b,b\}=2 \; ,\qquad
 [x, \; b]= 0 = [\partial, \; b] \; ,
\end{equation}
 where $\{b,b'\}\equiv b\cdot b' + b'\cdot b$  denotes the anticommutator. To obtain (\ref{eq:ib1}) from (\ref{suposc}) and (\ref{suposc1}), we
 fix there $\epsilon = -1$ and specify the metric matrix as
 $$
 \bar{\varepsilon}^{ab} = {\footnotesize  \left(\!\!
 \begin{array}{ccc}
 0 & -1 & 0 \\
  1 & 0 & 0 \\
   0 & 0 & 2
 \end{array} \!\! \right)} \;\;\; \Rightarrow \;\;\;
 \varepsilon_{ab} = {\footnotesize  \left(\!\!
 \begin{array}{ccc}
 0 & 1 & 0 \\
  -1 & 0 & 0 \\
   0 & 0 & 1/2
 \end{array} \!\! \right)} \; .
 $$
 Note that the fermionic variable $b$ can be understood in the matrix
 representation as a single Pauli matrix (say  $\tau^3$).
 To define the operator $z$ in (\ref{eq:Rh2}) we need two copies
 of super-oscillator algebras ${\cal A}_1$
 and  ${\cal A}_2$ with the generators $c_1^a = (x_1,\partial_1,b_1)$
  and $c_2^a = (x_2,\partial_2,b_2)$ which act in
 two different spaces ${\cal{V}}_1$ and ${\cal{V}}_2$.
 Then, the invariant operator $z$ in (\ref{eq:Rh2})
 looks like
\begin{equation} \label{eq:zab}
z=\sigma c_1^a \, c_{2a}=\sigma \varepsilon_{ab} c_1^a \, c_{2}^{b} =
 \sigma(x_1\partial_2-x_2\partial_1)+\frac{\sigma}{2} b_1b_2
 \equiv {\sf x} + {\sf b} \, ,
\end{equation}
$$
{\sf x}=\sigma(x_1\partial_2-x_2\partial_1),\qquad
{\sf b}=\frac{\sigma}{2} \; b_1b_2,
\qquad\sigma=\pm i,
$$
where $b_i$ satisfy $b_i^2=1$ in view of (\ref{eq:ib1})
 and anticommute $b_1b_2=- b_2b_1$ in order to ensure (\ref{eq:c3}).
The characteristic equation for the fermionic part of $z$:
 \begin{equation} \label{eq:char1}
{\sf b}^2=\frac{\sigma^2}{4} b_1b_2b_1b_2=
-\frac{1}{4}  b_1b_2b_1b_2=\frac{1}{4} (b_1)^2(b_2)^2=\frac{1}{4} \; ,
 \end{equation}
(here we took into account (\ref{eq:ib1}))
 allows one to introduce the projection operators:
\begin{equation} \label{eq:po1}
P_{\pm\frac12}=\frac12\pm {\sf b},\qquad P_i \cdot P_j=
\delta_{ij}P_i,\quad i,j=\pm\frac12,\qquad
P_{+\frac12}+P_{-\frac12}=1 \, .
\end{equation}
Now any function $f$ of ${\sf b}$ can be decomposed in these
projectors
\begin{equation} \label{eq:bpo1}
{\sf b} \cdot P_{\pm \frac12} = \pm \frac12 \, P_{\pm \frac12} \;\;\;
\Rightarrow \;\;\;
f({\sf b})=f({\sf b})\cdot (P_{+\frac12}+ P_{-\frac12})
=f\big(1/2\big) \, P_{+\frac12}+
f\big(-1/2\big) \, P_{-\frac12} \, .
\end{equation}
Accordingly, the $R$-operator (\ref{eq:R6}) can also be decomposed as:
\begin{equation} \label{eq:R12}
\hat{\mathcal{R}}_{12}(u|z)=\hat{\mathcal{R}}_{12}(u|{\sf x} + {\sf b})=
\big(\frac12+{\sf b}\big) \cdot \hat{\mathcal{R}}_{12}\big(u|{\sf x} + \frac12\big)
+\big(\frac12-{\sf b}\big) \cdot \hat{\mathcal{R}}_{12}\big(u|{\sf x} - \frac12\big) \; .
\end{equation}
and finally we have
\begin{equation} \label{eq:Rr7}
\hat{\mathcal{R}}^{osp(1|2)}_{12}(u|z)=r_+(u,{\sf x})\frac{\Gamma\big(
\frac 12({\sf x}+\frac32+u)\big)}{\Gamma\big(\frac12({\sf x}+\frac32-u)\big)}
 \cdot P_{+\frac12}+r_-(u,{\sf x})\frac{\Gamma\big(\frac12({\sf x}+\frac12+u)\big)}
{\Gamma\big(\frac12({\sf x}+\frac12-u)\big)} \cdot P_{-\frac12},
\end{equation}
where $r_\pm(u,{\sf x}) = r(u,{\sf x} \pm \frac12)$ are  periodic functions
in ${\sf x}$, i.e., the general $osp(1|2)$-invariant $R$-operators consist of
two independent terms acting on two invariant subspaces,  corresponding
to eigenvalues $\pm\frac12$ of the fermionic part
${\sf b} \equiv \sigma b_1b_2$ of the invariant operator $z$. The
coefficients in the expansion (\ref{eq:Rr7}) in projectors
$P_{\pm\frac12}$ are the functions of the bosonic
part ${\sf x}=\sigma(x_1\partial_2-x_2\partial_1)$
 of the invariant operator $z$. These coefficients are nothing but the
 $R$-operators for the bosonic subalgebra
 $s\ell(2) \simeq sp(2) \subset osp(1|2)$.

\subsection{The case of $osp(2|2)$}
In this case, we have two bosonic $c^1=x$, $c^2=\partial$ and two
fermionic $c^3=b^1$, $c^4=b^2$, oscillators which we realize using
even and odd variables with the commutation relations (\ref{suposc}):
 \begin{equation} \label{eq:ib2}
 [x, \; \partial] = -1 \; ,\qquad \{b^\alpha,b^\beta\}=
 2 \delta^{\alpha\beta}\; ,\qquad
 [x, \; b^\alpha]= 0 = [\partial, \; b^\alpha] \; .
\end{equation}
Here again we fix $\epsilon = -1$ and
 $$
 \bar{\varepsilon}^{ab} = {\footnotesize  \left(\!\!
 \begin{array}{cccc}
 0 & -1 & 0 &0\\
  1 & 0 & 0 &0\\
   0 & 0 & 2 &0\\
      0 & 0 & 0 & 2
 \end{array} \!\! \right)} \;\;\; \Rightarrow \;\;\;
 \varepsilon_{ab} = {\footnotesize  \left(\!\!
 \begin{array}{cccc}
 0 & 1 & 0 &0\\
  -1 & 0 & 0 &0\\
   0 & 0 & 1/2&0\\
      0 & 0 & 0&1/2
 \end{array} \!\! \right)} \; .
 $$
 We introduce two super-oscillator algebras
${\cal A}_1$ and ${\cal A}_2$ with the generators $\{ c^a_1 \}$ and
$\{ c^a_2 \}$, respectively. The invariant operator (\ref{eq:Rh2}) is
\begin{equation} \label{eq:zab2}
z=\sigma\e_{ab}c_1^ac_2^b
=\sigma(x_1\partial_2-x_2\partial_1)+ \frac{\sigma}{2} \;
b^\a_1 \, b^\a_2 \equiv\sf x+ \sf b,
\end{equation}
where $\sigma=\pm i$.

 The fermionic oscillators $b^\alpha_i \in
{\cal A}_i$ with commutation relations (\ref{eq:ib2}) and (\ref{eq:c4})
generate the 4-dimensional Clifford algebra. It is well known
(see, e.g., \cite{IsRub}) that the generators of this Clifford algebra
can be realized in terms the Pauli matrices $\tau^\a$:
$$
b^1_1 = \tau^1  \otimes I_2 \, , \;\;\;
b^2_1 = \tau^2  \otimes I_2 \, ,\;\;\;
b^1_2 = \tau^3 \otimes \tau^1  \, , \;\;\;
b^2_2 = \tau^3 \otimes \tau^2  \, ,
$$
where $I_2$ is the unit $(2 \times 2)$ matrix.

The characteristic equation for the fermionic part
${\sf b} = \frac{\sigma}{2} \, b^\a_1 \, b^\a_2$
of the operator (\ref{eq:zab2}) is
\begin{equation} \label{eq:char2}
\sf b(\sf b^2-1)=0.
\end{equation}
The invariant subspaces spanned by the eigenvectors
corresponding to eigenvalues $0$, $\pm1$ of ${\sf b}$
are extracted by the projectors:
\begin{equation} \label{eq:proj2}
P_0=1-{\sf b}^2,\;\;\; P_{+1}=\frac12({\sf b}^2+{\sf b}),\;\;\;
 P_{-1} =\frac12({\sf b}^2-{\sf b}),\qquad
P_0 + P_{+1} + P_{-1} = 1 \; .
\end{equation}
The $R$-operator is decomposed as follows:
\begin{equation} \label{eq:R22}
\begin{array}{c}
\hat{\mathcal{R}}_{12}(u|z)=\hat{\mathcal{R}}_{12}(u|{\sf x}+{\sf b})
= \\ [0.2cm]
=
\hat{\mathcal{R}}_{12}(u|{\sf x +b})(P_0+P_{+1}+P_{-1})
=\sum\limits_{\ell=0,\pm1}\hat{\mathcal{R}}_{12}(u|{\sf x}+\ell)P_\ell .
\end{array}
\end{equation}
Then (\ref{eq:R6}) implies that the spinor-spinor $R$-operator
invariant with respect to $osp(2|2)$ supersymmetry has the form
\begin{equation} \label{eq:R27}
\hat{\mathcal{R}}_{12}^{osp(2|2)}(u|z)=\sum_{\ell=0,\pm1}r(u|{\sf x}+
\ell)\frac{\Gamma\big(\frac12({\sf x}+\ell+1+u)\big)}{\Gamma\big(
\frac12({\sf x}+\ell+1-u)\big)}P_\ell,\qquad r(u|z+2)=r(u|z).
\end{equation}
Note that in view of the periodicity condition
$r(u|{\sf x}-1)=r(u|{\sf x}+1)$ one can rewrite
(\ref{eq:R27}) as follows:
\begin{equation} \label{eq:R28}
\hat{\mathcal{R}}_{12}^{osp(2|2)}(u|z)=r(u|{\sf x})
\frac{\Gamma\big(\frac12({\sf x}+1+u)\big)}{\Gamma\big(
\frac12({\sf x}+1-u)\big)}(1-{\sf b}^2)+\frac12r(u|{\sf x}+1)
\frac{\Gamma\big(\frac12({\sf x}+u)\big)}{\Gamma\big(
\frac12({\sf x}-u)+1\big)}({\sf x}{\sf b}^2+u{\sf b}).
\end{equation}
In the pure bosonic case of the
orthogonal algebras $so(2k)$, the general solution for the
$\hat{\mathcal{R}}$-operator splits
into two independent solutions corresponding to two nonequivalent
chiral left and right representations (see \cite{CDI2}, \cite{IKK15},
\cite{KaKi}). This does not happen here in the super-symmetric case,
where the
even and odd functions of ${\sf b}$ are not separated,
due to the dependence on the bosonic operator ${\sf x}$
in the coefficients $r(u|{\sf x}+\ell)$ which
mixes the chiral representations of $so(2k)$.

\subsection{The case of $osp(n|2)$}

This case is a generalization of the examples considered in the
previous subsections (for $n=1$ and $n=2$ we respectively
reproduce the results for the $osp(1|2)$ and $osp(2|2)$ algebras).
 Consider the super-oscillator algebra ${\cal A}$
with two bosonic $c^1=x$, $c^2=\partial$
and $n$ fermionic generators
$c^{2+\alpha}= b^\alpha$ $(\alpha=1,...,n)$
with the commutation relations
(\ref{suposc}):
 \begin{equation} \label{eq:ib3}
 [x, \; \partial] = -1 \; ,\qquad \{b^\alpha,b^\beta\}=
 2 \delta^{\alpha\beta}\; ,\qquad
 [x, \; b^\alpha]= 0 = [\partial, \; b^\alpha] \; ,
\end{equation}
where the fermionic elements $b^\alpha$ are the generators of the
$n$-dimensional Clifford algebra. This corresponds to the choice of
 the parameter $\epsilon=-1$
and the metric tensor in the block form
\begin{equation} \label{eq:eps}
 \bar{\varepsilon}^{ab} = {\footnotesize  \left(\!\!
 \begin{array}{ccc}
 0 & -1 & 0 \\
  1 & 0 & 0 \\
   0 & 0 & 2 \, I_n
 \end{array} \!\! \right)} \;\;\; \Rightarrow \;\;\;
 \varepsilon_{ab} = {\footnotesize  \left(\!\!
 \begin{array}{ccc}
 0 & 1 & 0 \\
  -1 & 0 & 0 \\
   0 & 0 & \frac{1}{2} I_n
 \end{array} \!\! \right)} \;
 \end{equation}
where $I_n$ stands for the $n \times n$ unit matrix.
The invariant operator $z \in {\cal A} \otimes {\cal A}$ is
\begin{equation} \label{eq:zzn}
z=\sigma \varepsilon_{ab} c^a_1c^b_2=\sigma(x_1\partial_2-x_2\partial_1)+
\frac\sigma2 \, b^\alpha_1 b^\alpha_2
\equiv{\sf x}+{\sf b} \, ,
\end{equation}
 where $\{ c^a_1 \}$ and $\{ c^a_2 \}$ are the generators of
 the first and second factor in ${\cal A} \otimes {\cal A}$,
 and the cross-commutation relations of
 $\{ c^a_1 \}$ and $\{ c^a_2 \}$ are given in (\ref{eq:c4}).
The characteristic equation for the fermionic
part ${\sf b} = \frac\sigma2 \, b^\alpha_1 b^\alpha_2$ of the invariant
$z$ has the order $n+1$ (cf. (\ref{eq:char1}) and (\ref{eq:char2})):
 \begin{equation} \label{eq:char3}
 \prod_{m=0}^n \big( {\sf b} -m + \frac{n}{2}\big)=0 \; .
 \end{equation}
One can prove (\ref{eq:char3}) by noticing that the operator
${\sf b}$ is represented as
${\sf b} = \bar{z}^{\,\alpha} z^\alpha -n/2$,
where (see (\ref{cpm5})) $\bar{z}^{\,\alpha} \equiv
\frac{1}{2}(b^\alpha_1 - \sigma b^\alpha_2)=c^{2+\alpha}_-$ and
$z^\alpha \equiv \frac{1}{2}(b^\alpha_1 + \sigma b^\alpha_2)
=c^{2+\alpha}_+$ are
respectively the creation and annihilation fermionic operators in the
Fock space ${\cal F}$ which is created from the vacuum $|0\rangle$:
$z^\alpha |0\rangle = 0$ $(\forall \alpha)$.
Then the operator in the left-hand side of (\ref{eq:char3})
is equal to zero since it is zero on all basis vectors
$\bar{z}^{\,\alpha_1}\cdots \bar{z}^{\,\alpha_m}|0\rangle \in {\cal F}$
 (here $1 \leq \alpha_1 < ...< \alpha_m \leq n$ and $m \leq n$)
which are the eigenvectors of ${\sf b}$  with eigenvalues
$(m-\frac{n}{2})$.

The projectors $P_{\ell}$ on
invariant subspaces in ${\cal F}$ spanned by the eigenvectors
of ${\sf b}$ corresponding to eigenvalues $(m-\frac{n}{2})\equiv \ell$,
where $\ell=-\frac{n}{2},-\frac{n}{2}+1,...,\frac{n}{2}$,
are immediately obtained from (\ref{eq:char3}):
 \begin{equation} \label{proj3}
 P_{\ell} = \prod_{\stackrel{m \neq \ell}{m=-n/2}}^{n/2}
 \frac{ {\sf b} -m }{\ell - m}  \; , \;\;\;\;\;\;
 {\sf b} \cdot P_\ell   = \ell \, P_\ell \; , \;\;\;\;\;\;
 \sum_{\ell=-n/2}^{n/2} P_\ell = 1 \; .
 \end{equation}

 \subsubsection*{The case of even $n=2k$}
We see that eigenvalues of ${\sf b}$ are integer (or half-integer)
when the number $n$ is even (or odd). Thus, for the case
$osp(n|2)=osp(2k|2)$, when $n=2k$ is even, the expansion of the
 solution (\ref{eq:R6}) goes over integer eigenvalues
\begin{equation} \label{eq:R77}
 \begin{array}{c}
\hat{\mathcal{R}}^{osp(2k|2)}_{12}(u|z)=
\hat{\mathcal{R}}^{osp(2k|2)}_{12}(u|{\sf x + b}) =
 \sum\limits_{\ell=-k}^k\hat{\mathcal{R}}^{osp(2k|2)}_{12}
 (u|{\sf x}+\ell)P_{\ell},
\end{array}
\end{equation}
and it implies
\begin{equation} \label{eq:rr2k}
\hat{\mathcal{R}}^{osp(2k|2)}_{12}(u|z)=\sum_{\ell=-k}^kr(u|{\sf x}+
\ell)\frac{\Gamma\big(\frac12({\sf x}+\ell+1+u)\big)}{\Gamma\big(
\frac12({\sf x}+\ell+1-u)\big)}P_\ell,\qquad r(u|z+2)=r(u|z).
\end{equation}

 \subsubsection*{The case of odd $n=(2k+1)$\label{osp2kn}}

 For the case $osp(n|2)=osp(2k+1|2)$, when $n=2k+1$ is odd, the
 expansion of the solution (\ref{eq:R6}) goes over half-integer
 eigenvalues of ${\sf b}$: $-\frac{2k+1}2,-\frac{2k-1}2,\ldots,-\frac12,\frac12,\frac32,\ldots,
\frac{2k+1}2$, and we have the expansion
\begin{equation} \label{eq:bbn2}
\hat{\mathcal{R}}^{osp(2k+1|2)}_{12}(u|z)=
\hat{\mathcal{R}}^{osp(2k+1|2)}_{12}(u|{\sf x+b})
=\sum_{\ell=-k-\frac12}^{k+
\frac12}\hat{\mathcal{R}}^{osp(2k+1|2)}_{12}(u|{\sf x}+\ell)P_{\ell},
\end{equation}
which for solution (\ref{eq:R6}) implies
\begin{equation} \label{eq:rr2k1}
\hat{\mathcal{R}}^{osp(2k+1|2)}_{12}(u|z)=\sum_{\ell=-k-\frac12}^{k+
\frac12}r(u|{\sf x}+\ell)\frac{\Gamma\big(\frac12({\sf x}+\ell+1+u)\big)}
{\Gamma\big(\frac12({\sf x}+\ell+1-u)\big)}P_\ell \; ,
\end{equation}
where the periodic function $r(u|z+2)=r(u|z)$ normalizes the solution.

\subsection{The case of $osp(n|2m)$}
We consider the $osp(n|2m)$ invariant super-oscillator
algebra which is realized in terms
of $m$ pairs of the bosonic oscillators $c^{j}=x^j$,
$c^{m+j}=\partial^j$, $j=1,\ldots,m$ and $n$ fermionic
oscillators $c^{2m +\alpha} = b^\alpha$, $\alpha=1,2,\ldots,n$,
with the commutation relations  (\ref{eq:ib4})
 deduced from (\ref{suposc}) with the
 choice of the parameter $\epsilon=-1$ and metric tensor
 (\ref{eq:eps5}). In this case, the invariant operator
$z \in {\cal A} \otimes {\cal A}$ defined in (\ref{eq:Rh2}) is
\begin{equation} \label{eq:zzz}
z=\sigma \varepsilon_{ab} c^a_1c^b_2 =
\sigma\sum_{j=1}^m(x^{j}_1\partial^{j}_2
 -x^{j}_2\partial^{j}_1)
 + \frac\sigma2 \sum_{\alpha=1}^n b_1^\alpha \, b_2^\alpha
 \equiv{\sf x}+{\sf b}.
\end{equation}
 Here the operator ${\sf b}$ is the same as in the previous examples
 of Section {\bf \ref{osp2kn}}. Thus, the $R$ operator (\ref{eq:R6})
 in the case of the algebra $osp(n|2m)$
 is expanded over the projection operators $P_\ell$ like
in the case of $osp(n|2)$, and the final expression
for $\hat{\mathcal{R}}^{osp(2m|n)}_{12}(u|z)$ will be given by
(\ref{eq:rr2k}) or (\ref{eq:rr2k1}):
\begin{equation} \label{eq:rm2k}\!\!
\hat{\mathcal{R}}^{osp(n|2m)}_{12}(u|z)=\sum_{\ell\in\Omega_n}
r(u|{\sf x}+\ell)\frac{\Gamma\big(\frac12({\sf x}+\ell+1+u)\big)}
{\Gamma\big(\frac12({\sf x}+\ell+1-u)\big)}P_\ell \; ,
\end{equation}
where
$$
{\sf x}=\sigma\sum\limits_{j=1}^m(x^{j}_1\partial^{j}_2
 -x^{j}_2\partial^{j}_1) \; , \;\;\;\;\; r(u|z) = r(u|z+2) \; ,
 $$
the projectors $P_\ell$ are defined in (\ref{proj3}) and
\begin{equation} \label{eq:nnn}
\Omega_n=\left\{\ba{cc}\!\!\{-k,\;1-k,\;\ldots,k-1,k\},\;\;\quad n=2k,\\
\{-k-\frac12,\;\ldots,k+\frac12\},\qquad n=2k+1 \ea \right.\qquad\qquad
k\in{\mathbb{N}} \; .
\end{equation}

\section{The relation between two approaches\label{relapp}}
\setcounter{equation}0

In this section, we give a more direct and elegant derivation of the
$R$ matrix solution (\ref{anzR}), (\ref{Rrec1}), that does not require
the introduction of additional auxiliary variables (as it was done in
\cite{FIKK}) and is based only on using
  the generating function\footnote{However
we stress that generating
 function (\ref{rf3}) is obtained by using of the recurrence
 relation (\ref{rri}) while the latter is derived in the Appendix
 {\bf \ref{symprod}} by means of auxiliary variables.}
(\ref{rf}), (\ref{rf3}) of the invariant operators $\tilde{I}_k$.
In addition, this derivation partially explains the
 relationship between the two types of solutions (\ref{anzR}),
 (\ref{Rrec1}) and (\ref{eq:R6}) for the $R$ operator.

Now we clarify the relation of the Shankar-Witten form of the $R$ operator
(\ref{anzR}), (\ref{Rrec1}) and Faddeev-Takhtajan-Tarasov type $R$ operator
 given by the ratio of two Euler Gamma-functions in (\ref{eq:R6}). First,
we write (\ref{anzR}) in the form
\be\lb{chr}
\hat{\mathcal{R}}(z)=
\sum_{k=0}^\infty\frac{\tilde{r}_k(u)}{k!}\tilde{I}_k(z),
\ee
where $\tilde{r}_k(u) = (-\sigma)^k \, r_k(u)$ and $\tilde{I}_k =
\sigma^k \, I_k$. Recall that $\tilde{I}_k$ are the Hermitian
invariants introduced in the proof of Proposition {\bf \ref{reqinv}}.

\begin{proposition}
The $R$ operator (\ref{chr}) obeys
(\ref{eq:RLL4}) or equivalently the finite-difference equation
 (\ref{eq:R7}):
\be\lb{fid}
W\equiv(z-u) \, \hat{\mathcal{R}}(z+1)
 -(z+u)\, \hat{\mathcal{R}}(z-1)=0 \; ,
\ee
(which was used to find the second solution (\ref{eq:R6}))
 if the coefficients $\tilde{r}_k(u)$ satisfy
 \be\lb{chr1}
 \qquad \qquad \tilde{r}_{k+2}(u)=-
 \frac{4(u-k)}{k+2+u-\o}\; \tilde{r}_k(u) \, ,
\ee
that in terms of $r_k(u)$ is written as  (\ref{Rrec1}).
\end{proposition}
\noindent
{\bf Proof.}
One can write (\ref{fid}) as
\be\lb{w}
W=\sum_{k=0}^\infty
\frac{\tilde{r}_k(u)}{k!}\Big((z-u)\tilde{I}_k(z+1)-(z+u)
\tilde{I}_k(z-1)\Big)
=\ee
$$
=\sum_{k=0}^\infty\frac{\tilde{r}_k(u)}{k!}\Big((z-u)\dd_x^kF(x|z+1)-
(z+u)\dd_x^kF(x|z-1)\Big)_{x=0} \; .
$$
We use the relation (\ref{rf2}) in the form:
$$
zF(x|z)=\left[\bigl(1-\frac{x^2}4\bigr)\dd_x
 +\frac{\o x}4\right] \, F(x|z),
$$
and obtain
\be\lb{w1}
W=\sum_{k=0}^\infty\frac{\tilde{r}_k(u)}{k!}\dd_x^k
 \Big((z+1-u-1)F(x|z+1)-
(z-1+u+1)F(x|z-1)\Big)_{x=0}=
\ee
$$
=\sum_{k=0}^\infty\frac{\tilde{r}_k(u)}{k!}\dd_x^k\Big([(1-\frac{x^2}4)\dd_x+
\frac{\o x}4-u-1]F(x|z+1)-
[(1-\frac{x^2}4)\dd_x+\frac{\o x}4+u+1]F(x|z-1)\Big)_{x=0},
$$
Then we use the equations
$$
F(x|z+1)=\frac{1+\frac x2}{1-\frac x2}F(x|z),\qquad\qquad
F(x|z-1)=\frac{1-\frac x2}{1+\frac x2}F(x|z),
$$
which follow from formula (\ref{rf3}) for the generating function
and write (\ref{w1}) in the form
\be\lb{w11}
W= \sum_{k=0}^\infty\frac{r_k(u)}{k!}\dd_x^k
 \left\{(1-\frac{x^2}4)x\dd_x
 -u -(u-\o+2)\frac{x^2}4\right\}
\Big(\frac{2F(x|z)}{1-\frac{x^2}4} \Big)_{x=0}.
\ee
Now we apply the identity $\dd^k_x\;x=x\dd^k_x+k\dd^{k-1}_x$ to move
derivatives $\dd_x$ in (\ref{w11}) to the right and obtain:
\be\lb{w2}
W =\sum_{k=0}^\infty\frac{\tilde{r}_k(u)}{k!}\left[ (k-u)\dd_x^k-
\frac{k+u- \o}4k(k-1)
\dd_x^{k-2}\right] \Big(\frac{2F(x|z)}{1-\frac{x^2}4} \Big)_{x=0} \; .
\ee
In the second term in square brackets we shift the summation
parameter $k \to k+2$ and deduce
$$
W =
\sum_{k=0}^\infty\frac1{k!}\Big((k-u)\tilde{r}_k(u)
-\frac{k+2+u-\o}4\tilde
{r}_{k+2}(u)\Big)\dd_x^k\Big(\frac{2F(x|z)}{1-\frac{x^2}4} \Big)_{x=0}=0.
$$
The resulting expression vanishes due to
(\ref{chr1}). Thus, we prove that the finite-difference
 equation (\ref{fid}) is valid if the coefficients
 $\tilde{r}_k(u)$ satisfy (\ref{chr1}).
\hfill \qed

\vspace{0.2cm}

\noindent
{\bf Remark 11.}
 We prove that both
 $R$ operators (\ref{chr}), (\ref{chr1}) and (\ref{eq:R6}) satisfy
 the same equation (\ref{fid}) and indeed
 obey the $RLL$ relations (\ref{RLL51}).
It is worth also to note that the differential operator in the curly brackets in
(\ref{w11}) coincides (up to change of variable $x = \sigma \lambda$)
with the differential operator in curly brackets presented in formula
(6.43) of our work \cite{FIKK}. This suggests to regard
  the generating function $F(x|z)(1-\frac{x^2}4)^{-1}$
  as a coherent state in the super-oscillator space.

\section*{Acknowledgments}

The authors would like to thank S.Derkachov for useful discussions and
comments. A.P.I. acknowledges the support of the Russian Science
Foundation, grant No. 19-11-00131. The work of D.K. was partially
supported by the Armenian State Committee of Science grant 18T-132
and by the Regional Training Network on Theoretical Physics sponsored
by Volkswagenstiftung Contract nr. 86 260.

\newpage
\appendix

\section{Properties of operators ${\cal P}$, ${\cal K}$ and relations
for matrix generators of Brauer algebra\label{PKprop}}
\setcounter{equation}0

We use here the concise matrix notation introduced
 in Sections {\bf \ref{supgr}, \ref{sec2}}
  (this convenient notation was proposed in \cite{FRT}).
   \marginpar{\bf AIf}
 The matrices (\ref{osp07}) satisfy the identities
$$
 {\mathcal{P}}_{12} = {\mathcal{P}}_{21} \; , \;\;\;
 {\mathcal{K}}_{12} =
 (-)^{12}{\mathcal{K}}_{21}  (-)^{12} \; , \;\;\;
 (-)^{1} {\mathcal{K}}_{12}  =
 (-)^{2} {\mathcal{K}}_{12} \; , \;\;\;
 {\mathcal{K}}_{12} (-)^{1}  =
  {\mathcal{K}}_{12} (-)^{2} \; ,
$$
\begin{equation}
 \label{ident00}
 \begin{array}{c}
{\mathcal{P}}_{12}{\mathcal{P}}_{12}= {\bf 1} \; , \;\;\;
{\mathcal{K}}_{12}{\mathcal{K}}_{12}=
\omega {\mathcal{K}}_{12} \; , \;\;\;
{\mathcal{K}}_{12}{\mathcal{P}}_{12} =
{\mathcal{P}}_{12}{\mathcal{K}}_{12} = \epsilon {\mathcal{K}}_{12},
\end{array}
 \end{equation}
 where $\omega = \epsilon(N-M)$, the operator
 $(-)^{12}$ is defined in (\ref{min12}) and we introduce the matrix
 $(-)^{i} = (-1)^{[a_i]} \delta^{a_i}_{b_i}$ of super-trace in the
 $i$-th factor ${\cal V}_{(N|M)}$ of the product ${\cal V}_{(N|M)}^{\otimes 2}$.
 Then, by making use definitions (\ref{PP12}) and
 (\ref{KK12}), we have
   \begin{equation}
 \label{ident16a}
  (-)^1 {\mathcal{P}}_{12} = {\mathcal{P}}_{12} (-)^2
   \; , \;\;\; (-)^{23} {\mathcal{P}}_{13} =
    {\mathcal{P}}_{13} (-)^{12} \; , \;\;\;
  {\mathcal{P}}_{13} (-)^{23} = (-)^{12} {\mathcal{P}}_{13} \; ,
   \end{equation}
 \begin{equation}
 \label{ident16}
 \begin{array}{c}
{\mathcal{P}}_{12}{\mathcal{P}}_{23}=
(-)^{12} {\mathcal{P}}_{13} (-)^{12} {\mathcal{P}}_{12}=
 {\mathcal{P}}_{23} (-)^{23} {\mathcal{P}}_{13} (-)^{23} =
 {\mathcal{P}}_{23} (-)^{12} {\mathcal{P}}_{13} (-)^{12} \; ,
  \\ [0.2cm]
 {\mathcal{P}}_{12}{\mathcal{K}}_{13}=
(-)^{12} {\mathcal{K}}_{23}(-)^{12} {\mathcal{P}}_{12} \; , \;\;\;
{\mathcal{K}}_{23} {\mathcal{P}}_{12} =
{\mathcal{P}}_{12}(-)^{12} {\mathcal{K}}_{13} (-)^{12} \; ,
\end{array}
 \end{equation}
 \begin{equation}
 \label{ident15a}
 \begin{array}{c}
 \epsilon {\mathcal{K}}_{12}{\mathcal{P}}_{31}=
{\mathcal{K}}_{12}(-)^{12} {\mathcal{K}}_{32} (-)^{12} \; , \;\;\;
\epsilon {\mathcal{P}}_{31}{\mathcal{K}}_{12}=
 (-)^{12} {\mathcal{K}}_{32} (-)^{12} {\mathcal{K}}_{12}\; ,
\end{array}
 \end{equation}
  \begin{equation}
 \label{ident15b}
  \begin{array}{c}
{\mathcal{K}}_{23}{\mathcal{K}}_{12} = \epsilon \, {\mathcal{K}}_{23}
 (-)^{12}{\mathcal{P}}_{13}(-)^{12}=
 \epsilon \,  (-)^{23}{\mathcal{P}}_{13}(-)^{23} {\mathcal{K}}_{12} \; ,
\end{array}
 \end{equation}
 \begin{equation}
 \label{ident15}
  \begin{array}{c}
{\mathcal{K}}_{31}{\mathcal{K}}_{12} = \epsilon \,
 (-)^{12}{\mathcal{P}}_{32}(-)^{12}{\mathcal{K}}_{12} =
 \epsilon \, {\mathcal{K}}_{31} (-)^{13}{\mathcal{P}}_{32}(-)^{13} \; .
\end{array}
 \end{equation}
 The mirror counterparts of identities (\ref{ident16}) --
 (\ref{ident15}) are also valid.
 The identities (\ref{ident16a}), (\ref{ident16}) follow from the representation
 (\ref{PP12}): ${\mathcal{P}}_{12}= (-)^{12} P_{12}=
 P_{12}(-)^{12}$, where $P_{12}$ is the usual permutation operator.
 The identities
 (\ref{ident15}) follow
 from the definitions (\ref{osp07}),
 (\ref{PP12}), (\ref{KK12}) of the operators ${\mathcal{P}}$
 and ${\mathcal{K}}$. We prove only the first equality
 in (\ref{ident15}) since the other identities
 in (\ref{ident15a}), (\ref{ident15b})
  and (\ref{ident15}) can be proved in the same way.
 We denote the incoming matrix indices by $a_1,a_2,a_3$ and
 the outcoming indices by
$c_1,c_2,c_3$ while summation indices are $b_i$ and $d_i$.
Then we have
 $$
 \begin{array}{c}
 ({\mathcal{K}}_{31}
 {\mathcal{K}}_{12})^{a_1 a_2 a_3}_{c_1 c_2 c_3} =
 \bar{\varepsilon}^{a_3 a_1} \varepsilon_{c_3 b_1}
 \bar{\varepsilon}^{b_1 a_2} \varepsilon_{c_1 c_2} =
 \bar{\varepsilon}^{a_3 a_1} \delta_{c_3}^{a_2} \varepsilon_{c_1 c_2}
  = \delta_{c_3}^{a_2} \delta^{a_3}_{b_2}
  \epsilon (-)^{[a_1][b_2]} \bar{\varepsilon}^{a_1b_2}
  \varepsilon_{c_1 c_2} = \\ [0.2cm]
  = \epsilon (-1)^{[a_2][a_3]}
  ({\mathcal{P}}_{23})^{a_2a_3}_{b_2c_3}
  (-1)^{[a_1][b_2]}
  ({\mathcal{K}}_{12})^{a_1b_2}_{\;\; c_1 c_2} =
  \epsilon \, \bigl({\mathcal{P}}_{23} (-)^{23} \, (-)^{12}  \,
  {\mathcal{K}}_{12}\bigr)^{a_1 a_2 a_3}_{\;\; c_1 c_2 c_3} \; ,
  \end{array}
 $$
 and in view of the relation $(-)^{23} {\mathcal{K}}_{12}=
 (-)^{13} {\mathcal{K}}_{12}$
 which follows from (\ref{SupMet2}) and obvious
 identity ${\mathcal{P}}_{23} \,(-)^{13} =
 (-)^{12} \, {\mathcal{P}}_{23}$
 we obtain the first equality in (\ref{ident15}).

By means of the relations (\ref{ident16}) -- (\ref{ident15})
 one can immediately check eqs. (\ref{osp08}),
 (\ref{osp09}) and also deduce
 \begin{equation}
 \label{ident01}
{\mathcal{P}}_{12}{\mathcal{P}}_{23}{\mathcal{P}}_{12}=
{\mathcal{P}}_{23}{\mathcal{P}}_{12}{\mathcal{P}}_{23}.
 \end{equation}
 \begin{equation}
 \label{ident02}
{\mathcal{K}}_{12}{\mathcal{K}}_{23}{\mathcal{K}}_{12}=
{\mathcal{K}}_{12}, \;\;\;\;
{\mathcal{K}}_{23}{\mathcal{K}}_{12}{\mathcal{K}}_{23}=
{\mathcal{K}}_{23},
\end{equation}
 \begin{equation}
 \label{ident05}
{\mathcal{P}}_{12}{\mathcal{K}}_{23}{\mathcal{K}}_{12}=
{\mathcal{P}}_{23}{\mathcal{K}}_{12} \; ,  \;\;\;
{\mathcal{K}}_{12}{\mathcal{K}}_{23}{\mathcal{P}}_{12}=
{\mathcal{K}}_{12}{\mathcal{P}}_{23},
\end{equation}
\begin{equation}
 \label{ident03}
{\mathcal{P}}_{23}{\mathcal{K}}_{12}{\mathcal{K}}_{23}=
{\mathcal{P}}_{12}{\mathcal{K}}_{23} \; , \;\;\;
{\mathcal{K}}_{23}{\mathcal{K}}_{12}{\mathcal{P}}_{23}=
{\mathcal{K}}_{23}  {\mathcal{P}}_{12} \; .
\end{equation}
The identity (\ref{ident01}) follows from
the relations in the first line of (\ref{ident16}).
We consider a few relations in (\ref{ident02}) -- (\ref{ident03})
 in detail. We start to prove the first relation in (\ref{ident02}):
$$
({\mathcal{K}}_{12}{\mathcal{K}}_{23}
{\mathcal{K}}_{12})^{a_1a_2a_3}_{c_1c_2c_3} =
\bar{\varepsilon}^{a_1a_2}\varepsilon_{b_1b_2}
\bar{\varepsilon}^{b_2a_3}\varepsilon_{d_2c_3}
\bar{\varepsilon}^{b_1d_2}\varepsilon_{c_1c_2} =
\bar{\varepsilon}^{a_1a_2}\delta_{b_1}^{a_3}\delta_{c_3}^{b_1}
\varepsilon_{c_1c_2} =
{\cal K}^{a_1a_2}_{\;\; c_1c_2}\delta_{c_3}^{a_3}\; .
$$
The second relation in (\ref{ident02})
can be proved in the same way. Then
we prove the first
 equation in (\ref{ident05}). For the left hand side
  of (\ref{ident05}) one has:
$$
\begin{array}{c}
({\mathcal{P}}_{12}{\mathcal{K}}_{23}
{\mathcal{K}}_{12})^{a_1a_2a_3}_{\;\; c_1c_2c_3} =
(-1)^{[a_1][a_2]}\delta^{a_1}_{b_2}\delta^{a_2}_{b_1}
\bar{\varepsilon}^{b_2a_3}\varepsilon_{d_2c_3}
\bar{\varepsilon}^{b_1d_2} \varepsilon_{c_1c_2}=
(-1)^{[a_1][a_2]}\bar{\varepsilon}^{a_1a_3}
\delta^{a_2}_{c_3} \varepsilon_{c_1c_2} = \\ [0.2cm]
 = \delta^{a_2}_{c_3}  \delta^{a_3}_{b_2}
 (-1)^{[a_1][c_3]} \bar{\varepsilon}^{a_1b_2} \varepsilon_{c_1c_2}
 = \delta^{a_2}_{c_3}  \delta^{a_3}_{b_2}
 (-1)^{[b_2][c_3]} \bar{\varepsilon}^{a_1b_2} \varepsilon_{c_1c_2} =
 ({\mathcal{P}}_{23}
 {\mathcal{K}}_{12})^{a_1a_2a_3}_{\;\; c_1c_2c_3} \; ,
\end{array}
$$
and similarly one deduces other relations
in (\ref{ident05}) and (\ref{ident03}).
From the identities (\ref{ident01}) -- (\ref{ident03})
 we also deduce the following relations:
 \begin{equation}
 \label{ident12}
 {\mathcal{K}}_{12}{\mathcal{P}}_{23}{\mathcal{K}}_{12}=
\epsilon{\mathcal{K}}_{12}, \;\;\;
{\mathcal{K}}_{23}{\mathcal{P}}_{12}{\mathcal{K}}_{23}=
\epsilon{\mathcal{K}}_{23}.
\end{equation}
 \begin{equation}
 \label{ident11}
{\mathcal{P}}_{12}{\mathcal{K}}_{23}{\mathcal{P}}_{12}=
{\mathcal{P}}_{23}{\mathcal{K}}_{12}{\mathcal{P}}_{23}.
 \end{equation}
 \begin{equation}
 \label{ident04}
{\mathcal{P}}_{12}{\mathcal{P}}_{23}{\mathcal{K}}_{12}=
{\mathcal{K}}_{23}{\mathcal{P}}_{12}{\mathcal{P}}_{23},
\;\;\;\;
{\mathcal{K}}_{12}{\mathcal{P}}_{23}{\mathcal{P}}_{12}=
{\mathcal{P}}_{23}{\mathcal{P}}_{12}{\mathcal{K}}_{23}.
\end{equation}
Indeed, if we act from the left on both sides of the first relation
in (\ref{ident05}) by ${\mathcal{K}}_{12}$
and use (\ref{ident00}), (\ref{ident02}) we obtain
the first relation in (\ref{ident12}). In the same way
one can deduce from the first relation in (\ref{ident03})
the second relation in (\ref{ident12}). Now we act
on both sides of (\ref{ident05}) by ${\mathcal{P}}_{23}$
 from the right and use the last equation in (\ref{ident03}).
  As a result, we arrive at identity (\ref{ident11}).
  Finally, relations (\ref{ident04})  trivially follow from eq.
  (\ref{ident11}).

At the end of this appendix, we stress that the identities (\ref{ident00}),
(\ref{ident01}) -- (\ref{ident03}) are the images
of the defining relations (\ref{defBrauer1}), (\ref{defBrauer2})
for the Brauer algebra in the
representation (\ref{Brauer}). The $R$-matrix
(\ref{RBrauer01}) is the image of the element (\ref{RBrauer})
 and the Yang-Baxter equation (\ref{eq:YBEbraid})
 is the image of the identity (\ref{YBEbr}).
Thus,  it follows from proposition
 {\bf \ref{Prop11}} that
 the $R$-matrix (\ref{RBrauer01}) is a solution
 of the braided version of the Yang-Baxter equation (\ref{eq:YBEbraid}).

\section{Supersymmetrized products of super-oscillators\label{symprod}}
\setcounter{equation}0

The product of $k$ super-oscillators is transformed under
 the action (\ref{transU}) of the group $Osp$ as follows:
 \begin{equation} \label{SS02}
 c^{a_1} c^{a_2} \cdots c^{a_k} \; \to \;
  \Delta^{(k-1)}(U)^{a_1 a_2...a_k}_{\;\; b_1 b_2...b_k}
 c^{b_1} c^{b_2} \cdots c^{b_k} \, ,
 \end{equation}
 where  $U \in Osp$, and the tensor product of $k$ defining
 representations of the group $Osp$ is given by the formula
 $$
 \Delta^{(k-1)}(U)^{a_1 a_2...a_k}_{\;\; b_1 b_2...b_k} =
  U^{a_1}_{\;\; b_1} (-1)^{b_1 a_2}
 U^{a_2}_{\;\; b_2} (-1)^{b_1 b_2} \cdots
  (-1)^{(\sum\limits_{j=1}^{k-1} b_j) a_k} U^{a_k}_{\;\; b_k}
  (-1)^{(\sum\limits_{j=1}^{k-1} b_j) b_k} \, ,
 $$
 or in the concise notation we have
 $$
 \Delta^{(k-1)}(U)_{12...k} = U_1 (-)^{[1][2]} U_2 (-)^{[1][2]}
 \cdots (-)^{[k] \sum\limits_{j=1}^{k-1} [j]} U_k
 (-)^{(\sum\limits_{j=1}^{k-1} [j])[k]} \, .
 $$
 One can check that any element $X \in B_k(\omega)$
  of the Brauer algebra (\ref{defBrauer1}), (\ref{defBrauer2})
  in the representation (\ref{Brauer}) commutes with
  the action of the $Osp$ supergroup
  \begin{equation} \label{SS03}
  \Delta^{(k-1)}(U)_{12...k} \cdot X = X \cdot
  \Delta^{(k-1)}(U)_{12...k} \; .
  \end{equation}

Define the super-symmetrized product of two super-oscillators
$c^a,c^b$ as
\begin{equation} \label{SS01}
c^{(a} c^{b)} \equiv \frac{1}{2}
\left( c^a c^b -\epsilon (-1)^{[a][b]} c^b c^a \right) = \frac{1}{2}
\left( c^a c^b -\epsilon \, {\cal P}^{ab}_{\; de}
\, c^d c^e \right) = (A_2)^{ab}_{\;\; de} c^d c^e \; ,
\end{equation}
where ${\cal P}$ is the super-permutation matrix (\ref{PP12})
and $(A_2)^{ab}_{\;\; de}$ is the antisymmetrizer
$A_2 = \frac{1}{2}(1 - s_1)$ in the representation (\ref{Brauer}).
 The direct generalization of (\ref{SS01}) to the
 super-symmetrized product of any number of super-oscillators is
 the following:
 \begin{equation} \label{cccA}
 c^{(a_1}c^{a_2}\cdots c^{a_k)} =
 (A_k)^{a_1 a_2...a_k}_{\;\; b_1 b_2...b_k} \;
  c^{b_1}c^{b_2} \cdots c^{b_k} \; ,
 \end{equation}
 where $A_k$ is the $k$-th rank antisymmetrizer in the representation
 (\ref{Brauer}). The antisymmetrizer $A_k$ can be defined
 via the recurrence relations (see, e.g., \cite{IsRub})
 $$
 \begin{array}{c}
 A_k = \frac{1}{k} A_{k-1} \big(1 - s_{k-1} +  s_{k-1} s_{k-2} - ...
 + (-1)^{k-1} s_{k-1}  \cdots s_{2} s_1 \big)= \\ [0.2cm]
 = \frac{1}{k} \big(1 - s_{k-1} +  s_{k-2} s_{k-1} - ...
 + (-1)^{k-1} s_{1} s_2 \cdots s_{k-1}\big) A_{k-1} \; ,
 \end{array}
 $$
  and after substituting here the recurrence relations
  for $A_{k-1}$, $A_{k-2}$ etc., we arrive at the factorised formula
 \begin{equation} \label{SS04}
 A_k = \frac{1}{k!} \big(1 - s_{k-1} +   ...
 + (-1)^{k-1} s_{1} s_2 \cdots s_{k-1} \big) \cdots
  \big(1 - s_{2} +  s_{1} s_{2} \big)
  \big(1 - s_{1} \big) \; .
 \end{equation}
 We stress that in view of the relation (\ref{SS03})
 the super-symmetrized product (\ref{cccA}) is transformed
 under the action of $Osp$ as a usual product (\ref{SS02}).

Note that upon opening the parentheses, the element (\ref{SS04})
equals the alternating sum of all $k!$ elements of the
permutation group $S_k$. Using this fact
one can give a more explicit formula for super-symmetrized product
 (\ref{cccA}) of a higher number of super-oscillators
\begin{equation} \label{eq:ccc}
c^{(a_1}c^{a_2}\ldots c^{a_k)}\equiv\frac1{k!}\sum_{\sigma\in S_k}(-
\epsilon)^{p(\sigma)}(-1)^{\hat\sigma}c^{a_{\sigma_1}}\ldots c^{a_{
\sigma_k}}=\frac1{k!}\partial_\kappa^{a_1} \cdots \partial^{a_k}_\kappa
(\kappa\cdot c)^k=
\end{equation}
$$
=\partial_\kappa^{a_1} \cdots \partial^{a_k}_\kappa
\;  \exp(\kappa_a \, c^a)|_{\kappa=0} \; ,
$$
where $p(\sigma)=0,1$ denotes the parity of the
permutation $\sigma$. Here we introduce (see \cite{FIKK}) auxiliary
super-vector $\kappa_a$ such that the derivatives $\partial^{a}_\kappa =
\frac{\partial}{\partial\kappa_{a}}$ satisfy
$$
\partial_\kappa^{a_1} c^{a_2} =
- \epsilon (-1)^{a_1a_2}c^{a_2}\partial_\kappa^{a_1} \, , \;\;\;
\partial_\kappa^{b} \, (\kappa_a \, c^a) = c^b +
(\kappa_a \, c^a) \, \partial_\kappa^{b} \, , \;\;\;
 \partial_\kappa^{a} \partial_\kappa^{b} =
 - \epsilon (-1)^{[a][b]}
 \partial_\kappa^{b} \partial_\kappa^{a}\, ,
$$
and (\ref{eq:ccc}) holds due to the Leibniz rule.

Now we explain the notation ${\hat{\sigma}}$ in (\ref{eq:ccc}).
Let $s_j\equiv \sigma_{j, j+1}$ be an elementary transposition
of the $j$-th and $(j+1)$-st site. For the transposition $s_j$ we
define $\hat{s_j}=[a_j][a_{j+1}]$. Then for a general
permutation $\sigma=
s_{j_1}s_{j_2}\dots s_{j_{k-1}}s_{j_k} \in S_k$,
 we have
\begin{equation}
\hat{\sigma}=[a_{j_k}][a_{j_k+1}]+
[a_{s_{j_k}(j_{k-1})}][a_{s_{j_k}(j_{k-1}+1)}]+\cdots+
[a_{s_{j_2}\cdots s_{j_k}(j_1)}][a_{s_{j_2}
\cdots s_{j_k}(j_1+1)}].
\end{equation}
As an example, according to the definition (\ref{eq:ccc}),
we have the relation useful in practice:
\begin{equation}
c^{(a_1}\cdots c^{a_i} \cdots c^{a_j}\cdots c^{a_k)} = (-\epsilon)
(-1)^{[a_i][a_j]+\sum\limits_{l=i+1}^{j-1}
([a_i]+[a_j])[a_l]} c^{(a_1}\cdots c^{a_j}
\cdots c^{a_i}\cdots c^{a_k)}.
\end{equation}
In eq. (\ref{invars}) we have defined the supersymmetric invariants $I_m$
Using this definition, the representation (\ref{eq:ccc}) and the definition
(\ref{eq:Rh2}) of $z$, we obtain the recurrence relation
\begin{equation} \label{eq:zim}
\begin{array}{c}
- \sigma \, I_m \, \cdot z = I_m \, I_1=
 \varepsilon_{a_1b_1}\ldots\varepsilon_{a_mb_m}\partial_{\kappa_1}
^{a_1}\ldots\partial_{\kappa_1}^{a_m}\partial_{\kappa_2}^{b_m}\ldots
\partial_{\kappa_2}^{b_1}\varepsilon_{ab} (\partial^a_{\kappa_1}+
\frac12\epsilon(-1)^a\kappa^a_1)(\partial^b_{\kappa_2}+\\ [0.2cm]
+\frac12\epsilon(-1)^b\kappa^b_2)e^{\kappa_1\cdot c_1+\kappa_2
\cdot c_2}|_{\kappa_i=0}=I_{m+1}+\frac14\varepsilon_{a_1b_1}\ldots
\varepsilon_{a_mb_m}\partial_{\kappa_1}^{a_1}\ldots\partial_{\kappa
_1}^{a_m}\partial_{\kappa_2}^{b_m}\ldots\partial_{\kappa_2}^{b_1}
\varepsilon_{ab}(-1)^a\times\\ [0.2cm]
\times \kappa^a_1\kappa^b_2e^{\kappa_1\cdot
c_1+\kappa_2\cdot c_2}|_{\kappa_i=0}=I_{m+1}-
\frac\epsilon4\sum_{i=1}^m\varepsilon_{a_1b_1}\ldots\e_{a_{i-1}b_{i-1}
}\e_{a_{i+1}b_{i+1}}\ldots\varepsilon_{a_mb_m}\partial_{\kappa_1}
^{a_1}\ldots\partial_{\kappa_1}^{a_{i-1}}\times\\ [0.2cm]\times
\partial_{\kappa_1}^{a_{i+1}}\ldots\partial_{\kappa_1}^{a_m}\big(\frac\o
\epsilon-\epsilon\kappa_1^d\dd_{\kappa_1,d}\big)\partial_{\kappa_2}
^{b_m}\ldots\partial_{\kappa_2}^{b_{i+1}}\partial_{\kappa_2}^{b_{i-1}}
\ldots\partial_{\kappa_2}^{b_1}e^{\kappa_1\cdot
c_1+\kappa_2\cdot c_2}|_{\kappa_i=0}
\\ [0.2cm]= I_{m+1}+\frac m4\big((m-1) - \omega\big) I_{m-1} \; .
\end{array}
\end{equation}
For  a proof of (\ref{eq:zim}) we refer to the papers \cite{IKK15}
(see analogous calculation in eq. (5.6) there) and \cite{FIKK}.
Taking into account the initial conditions
$I_0=1$ and $I_1=-\sigma z$, we deduce from (\ref{eq:zim}) that the
invariants $I_m$ are polynomials in $z$ of the order $m$.




\end{document}